\documentclass[12pt]{article}
\usepackage{amsmath}
\usepackage{setspace}
\usepackage{multirow}
\usepackage{adjustbox}
\usepackage{fancyhdr}
\usepackage{amssymb}
\usepackage{array}
\usepackage{adjustbox}
\usepackage{float}
\usepackage{cite}
\usepackage{longtable}
\usepackage{subfigure}
\usepackage{caption}
\usepackage{siunitx}
\usepackage{booktabs}
\newcommand{\ra}[1]{\renewcommand{\arraystretch}{#1}}
\usepackage[section]{placeins}
\usepackage[a4paper]{geometry}
\begin{document}
	\title{\textbf{Interfacial strain relief by periodic dislocation doublets emerging from rotationally related orientation relationships of Y\textsubscript{4}Zr\textsubscript{3}O\textsubscript{12} dispersions in ferrite matrix }}
	\author{\normalfont Sruthi Mohan*\textsuperscript{1,2}, Alphy George\textsuperscript{2,3}, R. Vijay\textsuperscript{4}, C. David\textsuperscript{1}, G. Amarendra\textsuperscript{2}}
	\date{\textsuperscript{1} Material Science Group, Indira Gandhi Centre for Atomic Research, Kalpakkam, India.\\
		\textsuperscript{2} Homi Bhabha National Institute, Mumbai, India\\
		\textsuperscript{3} Metallurgy and Materials Group, Indira Gandhi Centre for Atomic Research, Kalpakkam, India.\\
		\textsuperscript{4} International Advanced Research Centre for Powder Metallurgy and New Materials, Hyderabad, India\\
		\medskip*Corresponding author: \textit{sruthi@igcar.gov.in}}

\maketitle
\hrule
\section*{Abstract}
The trigonal/bcc orientation relationships (ORs) and their likelihood of occurrence are extensively studied using dispersed Y\textsubscript{4}Zr\textsubscript{3}O\textsubscript{12}($\delta$) nano-precipitates in bcc Fe ($\alpha$) matrix by means of transmission electron microscopy, image simulations and a crystallographic model. Two orientation relationships related by a rotation:$[1\bar20]_\delta||[111]_\alpha$ with $(21\bar2)_\delta||(\bar110)_\alpha$ and $[1\bar20]_\delta||[111]_\alpha$ with $(00\bar3)_\delta||(\bar110)_\alpha$, are established and periodic arrays of misfit dislocation doublets are identified at the strained interface in $(110)_\alpha$ for both ORs. Further eighteen energetically feasible ORs in Y\textsubscript{4}Zr\textsubscript{3}O\textsubscript{12}/bcc system are deduced by combining  stereographic projections, which include the two predominant ORs in this study and other ORs in literature. The orientation relationship which generates interface with a minimum number of dislocation doublets is the most frequent. 
\medskip
\medskip
\medskip

\par \noindent\textbf{Keywords}: Orientation relationships, Misfit dislocations, Interfacial strain, Oxide dispersion strengthened steel 
\newpage
\section{Main}
Nucleation, growth, morphology and point defect interactions of secondary phases in a material is determined by the nature of interfaces resulting from the orientation relationships of crystal systems involved, which are usually closely connected by small relative rotations fixed by crystal symmetries\cite{dahmen1982orientation}. The interfaces in various precipitate/matrix systems including the ones resulting from the Burger's\cite{burgers1934process}, Pitsche-Schrader\cite{rong1984crystallography} and Potter's\cite{rong1984crystallography,potter1973structure} orientation relationships in bcc/hcp systems and Bain, Nishiyama-Wassermann (N-W), Kurdjumov-Sachs (K-S), Greningerp-Troiano (G-T) and Pitsch\cite{pitsch1959martensite} orientation relationships in bcc/fcc systems are well studied with abundant experimental data\cite{ray1990transformation} and have been successfully reproduced by the crystallographic models, edge to edge model\cite{kelly2006edge,zhang2005crystallographic,zhang2005edge}, structural ledge model\cite{rigsbee1979computer, hall1972structure}, invariant line model\cite{luo1987invariant, howe1992comparison,dahmen1984invariant} and O-lattice theory\cite{balluffi1982csl,howe1992comparison}. 
\par Dispersion strengthened alloys reinforced by homogeneous distributions of nano-sized oxide phases are rich in complex interfaces involving crystal structures of various symmetries. Studies of these interfaces by extending existing crystallographic models are crucial in forecasting the strength, hardness, thermal and irradiation stability of dispersion strengthened materials for deployment in various industries. Oxide Dispersion strengthened (ODS) steels with cubic (Y\textsubscript{2}O\textsubscript{3}\cite{ribis2012interfacial,klimiankou2004tem}, Y\textsubscript{2}Hf\textsubscript{2}O\textsubscript{7}\cite{oka2013morphology}, Y\textsubscript{2}Ti\textsubscript{2}O\textsubscript{7})\cite{ribis2012interfacial,hirata2011atomic}, monoclinic (Y\textsubscript{4}Al\textsubscript{2}O\textsubscript{9}\cite{hsiung2010formation}, YTaO\textsubscript{4}\cite{oka2013morphology}), hexagonal (Y\textsubscript{2}TiO\textsubscript{5}\cite{dou2019morphology}), orthorhombic (YTiO\textsubscript{3}\cite{dou2019crystal}, YAlO\textsubscript{3})
and trigonal (Y\textsubscript{4}Zr\textsubscript{3}O\textsubscript{12}\cite{dou2014tem}) dispersions have gained wide recognition as potential candidates for nuclear reactor applications, where these oxide phases provide high temperature strength and radiation resistance by pinning the dislocations, trapping irradiation induced point defects and limiting the grain growth\cite{odette2008recent,alinger2004development,alinger2008role,ukai1993alloying}. Although the ODS alloys are produced by high energy, non-equilibrium processes like ball milling, the matching of close packed or nearly close packed crystallographic planes and directions in reported literature indicate the possibility of structurally related and energetically favoured ORs existing with the matrix. However, comprehensive studies of orientation relationships in ODS systems are confined to a few cubic and monoclinic precipitates. 
\par Semi-coherent interfaces with orientations $[2\bar10]_\delta||[001]_\alpha$; $(12\bar4)_\delta||(\bar110)_\alpha$; $(122)_\delta||(\bar1\bar10)_\alpha$ and $[102]_\delta||[111]_\alpha$; $(4\bar3\bar2)_\delta||(10\bar1)_\alpha$; $(\bar231)||(\bar211)$; $(\bar231)_\delta||(\bar211)_\alpha$ are reported between trigonal Y\textsubscript{4}Zr\textsubscript{3}O\textsubscript{12} and cubic (bcc) Fe\cite{dou2014tem}. The low symmetry of trigonal systems cause different zone axes families to give same diffraction spots in transmission electron microscopy examinations and hence the image simulations are essential to determine the exact OR. The main focus of this study is to rigorously search for trigonal/bcc orientation relationships in Y\textsubscript{4}Zr\textsubscript{3}O\textsubscript{12}/Fe system using phase contrast transmission electron microscopy and multislice image simulations and predict the rotational transformations relating the ORs by constructing suitable stereographic projections.
\section{Nature of dispersoids in Fe-14Cr-0.6Zr-0.3Y\textsubscript{2}O\textsubscript{3} ODS alloy}		
\par
\begin{figure}[!hbt]
	\centering
	\includegraphics[width=\linewidth]{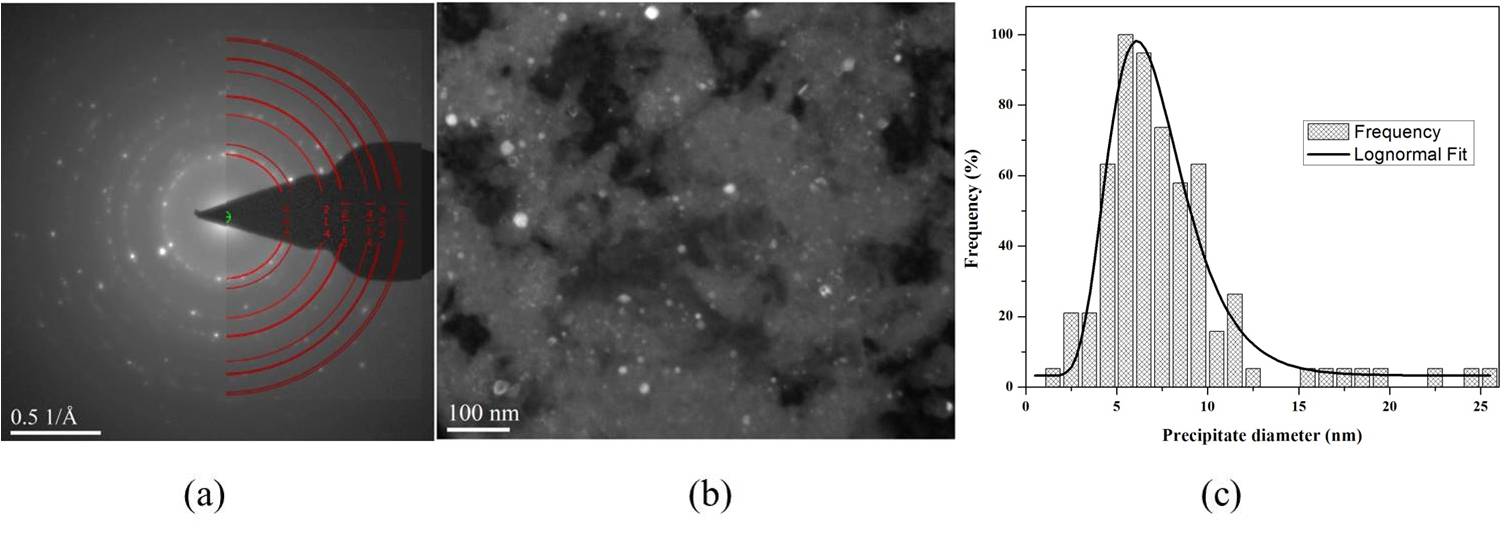}
	\caption{(a) SAED pattern of the extraction replica of Zr-ODS, which is indexed for trigonal Y\textsubscript{4}Zr\textsubscript{3}O\textsubscript{12} (ICDD : 01-077-0743) (b) Dark Field images of the carbon extraction replica of Zr-ODS steels (c) Size distribution of dispersoids with lognormal fit.}
	\label{BF-DF-Zr}
\end{figure}

 As documented in various literature, during the hot-consolidation, the dissolved minor alloying elements re-precipitate as homogeneously dispersed complex Y-Zr-O, Y-Ti-O or Y- Al-O oxides, depending on the alloy composition. Extraction replicas are made to obtain the structural information of oxides without ferromagnetic interference from the matrix. The dark field TEM images recorded from the carbon extraction replica specimen (Figure \ref{BF-DF-Zr}(b)), using the SAD  of Figure \ref{BF-DF-Zr}(a) show precipitates of size 1.6 -26 nm dispersed in the matrix with a mean diameter of 6 nm, as obtained from the log-normal fit of the size distribution (Figure \ref{BF-DF-Zr}(c)). 
\par The rings in SAD include dispersoids from multiple grains and are in exact agreement with trigonal Y\textsubscript{4}Zr\textsubscript{3}O\textsubscript{12}. The Y\textsubscript{4}Zr\textsubscript{3}O\textsubscript{12} is a fluorite related compound, known as $\delta$-phase, with $R\bar3$ space group. It is the most stable oxide in the Y-Zr-O system \cite{mohan2020ab} and can be represented in rhombohedral as well as hexagonal axes. The lattice parameters in hexagonal axes are: a=9.738 and c=9.115 (ICDD : 01-077-0743, \cite{red1991crystal}). In the 57 atom unitcell, 3a site is occupied by Zr atom, one set of 18f positions are shared by Zr and Y atoms with a Y-occupancy of 66.6\%. Two types of oxygen, generally denoted as O\textsubscript{I} and O\textsubscript{II} are distributed in the remaining two sets of 18f positions. The structure has two inherent oxygen vacancies at 6c sites aligned along the inversion triad, [111] direction.

\par According to Hsiung et al.\cite{hsiung2011hrtem}, the critical size for Y\textsubscript{4}Al\textsubscript{2}O\textsubscript{9} precipitates to posses definite structure and stoichiometry in $\alpha$-Fe matrix is 2 nm. Below this critical size, they were predominantly amorphous or disordered cluster domains. In the present study, even the precipitate of size as small as $\sim$ 1.6 nm can be resolved into the structure of Y\textsubscript{4}Zr\textsubscript{3}O\textsubscript{12} indicating that smaller precipitates can also have crystalline nature.

\begin{table}[!hbt]
	\centering
	\caption{Microstructural information of Zr ODS alloy. }
	\begin{center}
		\begin{tabular}{cp{1.5 cm}p{2 cm}p{2 cm}p{2 cm}}
			\toprule
			Alloy   & Grain size (nm) & Precipitate radius,$r$ (nm) & Number Density (/m\textsuperscript{3})  & Volume Fraction,$f$  \\ \midrule
			
			Zr - ODS & 234             & 3                         & 1.7$\times$10\textsuperscript{24}                               & 0.06            \\  
			\bottomrule
		\end{tabular}
		\newline
		\label{summary}
	\end{center}
\end{table}

\section{Predominant trigonal/bcc orientation relationship in Zr-ODS steels}
Detailed HRTEM investigation was carried out in the specimen prepared using twin jet electropolishing to determine the structure and orientation of the precipitates. The microstructural information obtained from these in-foil specimen are tabulated in Table \ref{summary}. The d-spacings and angles between the planes measured from the power spectrum of phase contrast image are compared with values obtained from the simulated diffraction pattern of different compounds in various zone axes. The major oxide phases considered for comparison are: Y\textsubscript{4}Zr\textsubscript{3}O\textsubscript{12}, ZrO\textsubscript{2}(cubic, monoclinic and tetragonal), Y\textsubscript{6}ZrO\textsubscript{11}, Cr\textsubscript{2}O\textsubscript{3}, Fe\textsubscript{2}O\textsubscript{3} and Y\textsubscript{2}O\textsubscript{3}(cubic, monoclinic).  The analysis confirm that all the precipitates in Zr -ODS have Y\textsubscript{4}Zr\textsubscript{3}O\textsubscript{12} structure.
\par
\begin{figure}[!hbt]
	\centering
	\subfigure[]{\includegraphics[width=0.32\linewidth]{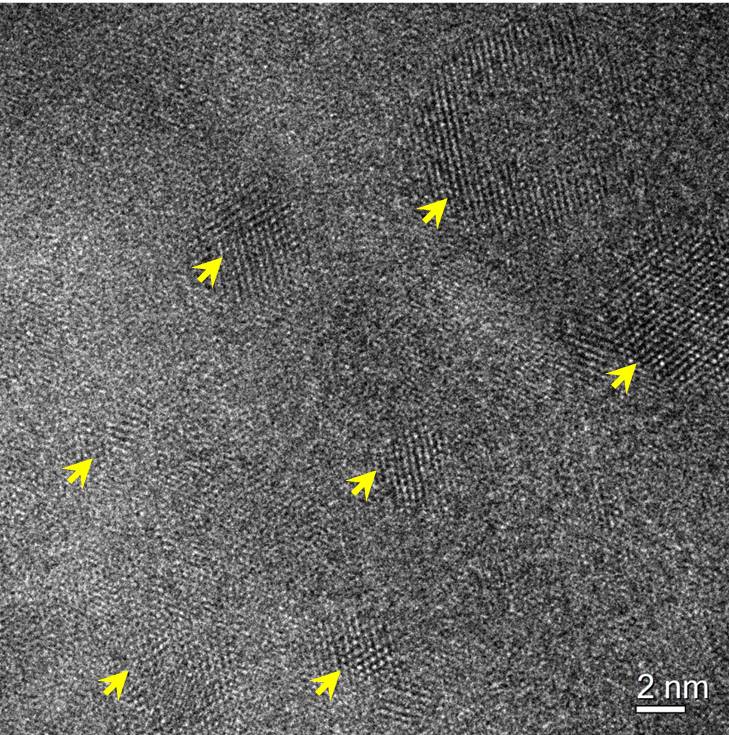}}
 	\subfigure[]{\includegraphics[width=0.32\linewidth]{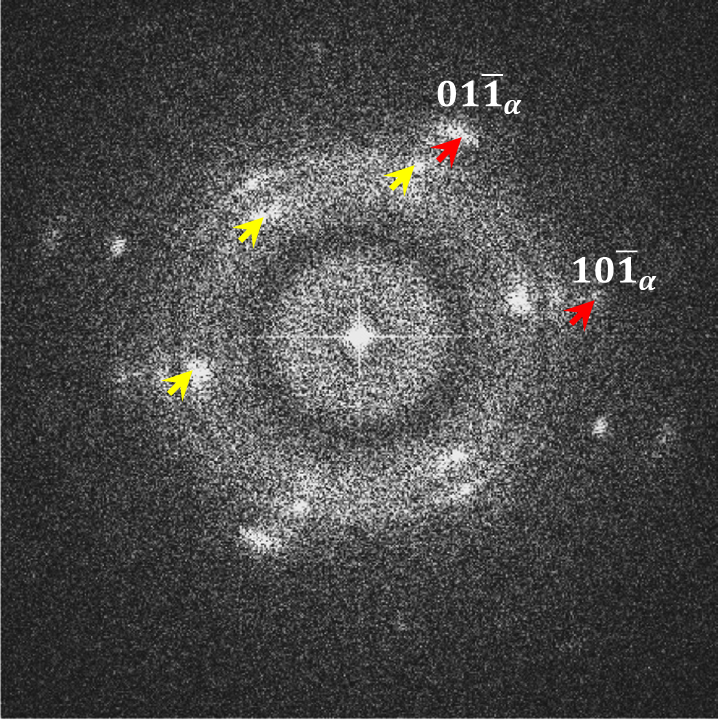}}
	\subfigure[]{\includegraphics[width=0.32\linewidth]{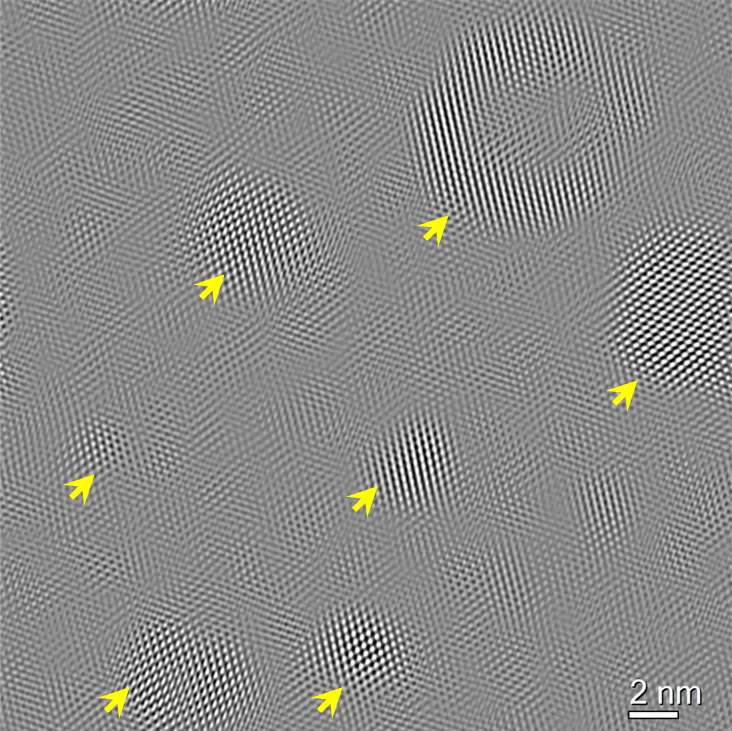}}
	\caption{ (a) HRTEM image showing distribution of dispersoids in the matrix of as prepared Zr-ODS steel. The dispersed precipitates in the micrograph have diameter ranging from 2 nm to 8 nm (b) Power spectrum of the micrograph in (a). Matrix spots are marked with red arrows and indexed for [111] zone axis of Fe. The diffraction spots corresponding to precipitates are marked with yellow arrows. (c) The image reconstructed using diffraction spots corresponding to precipitates. The nanoclusters are indicated by yellow arrows in (a) and (c).}
	\label{cluster}
\end{figure}

 A high resolution lattice image of the Zr ODS with a distribution of dispersoids in size range  2 nm to 8 nm (indicated by arrows) can be seen in Figure. \ref{cluster}(a). The power spectrum of the entire region is shown in \ref{cluster}(b) and the zone axis of the matrix ($\alpha$ - Fe) can be identified as [111]. The matrix spots are marked with red arrows, indexed and the precipitate spots are seen almost parallel to the matrix spots (marked in yellow arrows). The inverse Fourier transform image is constructed using precipitate spots alone and it is shown in \ref{cluster}(c). It can be seen that all the precipitates have the same orientation which is the predominant matrix-precipitate orientation relationship in [111] grains of Zr-ODS (designated as OR1).
\begin{figure}[!hbt]
	\centering
	\subfigure[]{\includegraphics[width=0.32\linewidth]{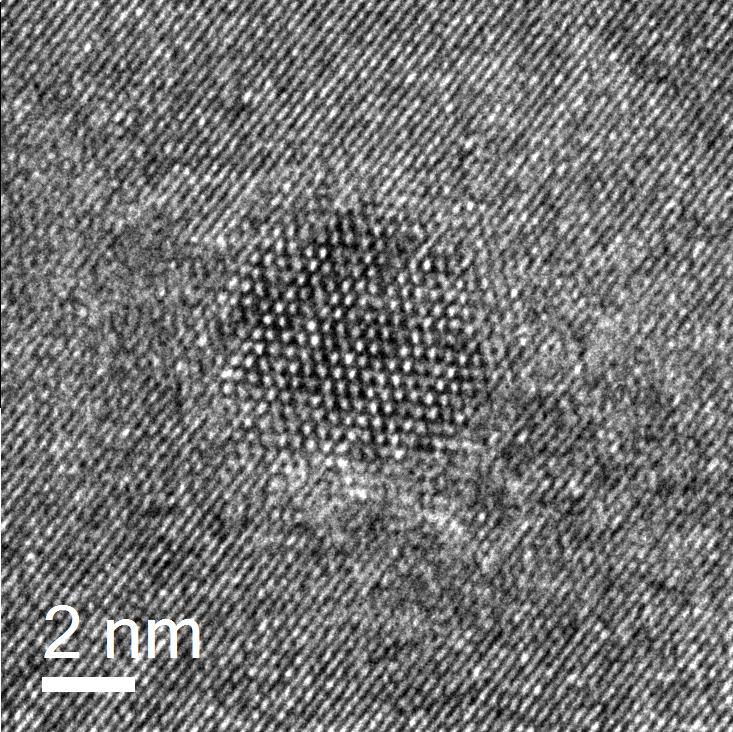}}
	\subfigure[]{\includegraphics[width=0.32\linewidth]{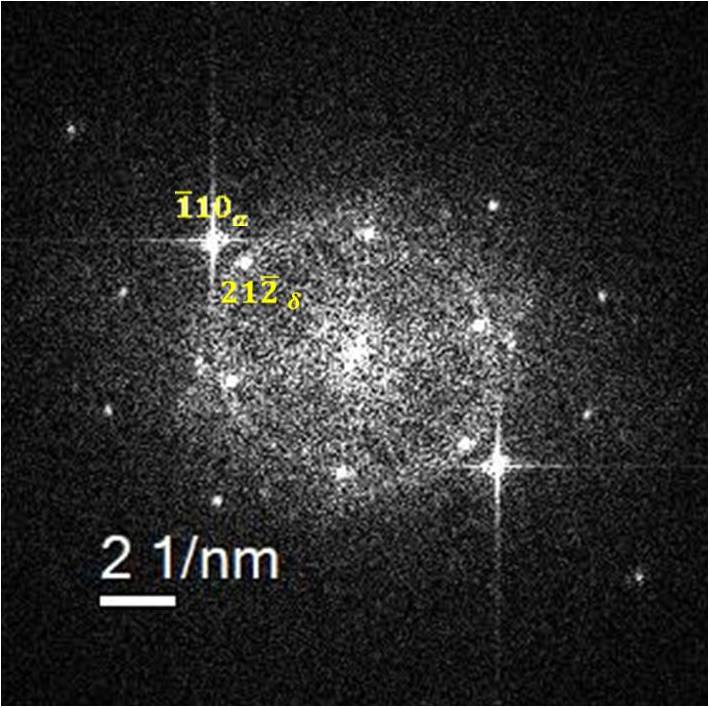}}
	\subfigure[]{\includegraphics[width=0.32\linewidth]{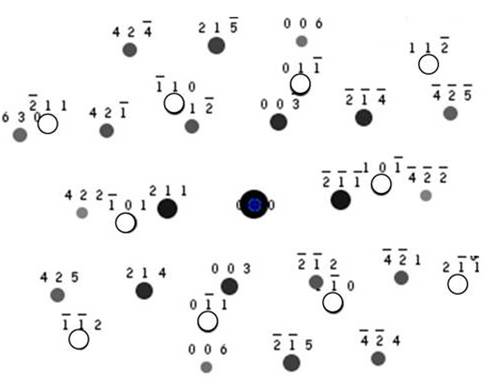}}
	\captionof{figure}{(a)HRTEM micrograph of a faceted Y\textsubscript{4}Zr\textsubscript{3}O\textsubscript{12} precipitate in Zr-ODS matrix. (b) Power spectrum of the image. The spots corresponding to the matrix are indexed with subscript Fe. (c) Simulated diffraction pattern corresponding to the orientation relationship $(10\bar1)_\alpha||(122)_\delta$ and $[111]_\alpha||[1\bar20]_\delta$ which can be derived from the power spectrum. The open circles indicate matrix-spots and filled circles indicate precipitate spots.}
	\label{2-10}
	\bigskip
	\captionof{table}{Inter-planar distances (d) and angles ($\alpha$) of the nanoparticle of Figure \ref{2-10}}
	\bigskip
	\begin{tabular}{cccccccc}
		\toprule
		\medskip
		&d(\AA),$\alpha(^\circ)$ &$d_1$ &$d_2$ &$d_3$ &$\alpha_{12}$ &$\alpha_{13}$ &$\alpha_{23}$\\\midrule
		\medskip
		&Measured &3.00  &1.85   &2.61  &35  &53 &88\\
		\medskip
		Particle &Planes	&$\left(003\right)$	&$\left(214\right)$		&$\left(21\bar2\right)$		&-	&-	&- \\ 
		\medskip
		&Calculated    &3.03	&1.85 	&2.61 	&35.56 	&55	&89 \\\midrule
		\medskip
		&Measured &2.04  &2.04 &2.04  &60  &120  &60\\
		\medskip
		Matrix  &Planes	&$\left(01\bar1\right)$	&$\left(10\bar1\right)$	&$\left(1\bar10\right)$	&-	&-	&- \\ 
		\medskip
		&Calculated    &2.02 &2.02	&2.02 	&60 	&120	&60\\\bottomrule
		\medskip
	\end{tabular}
	\label{tab:2-10}
\end{figure}

\par Magnified view of a dispersoid of the same orientation as in Figure \ref{cluster} is shown in Figure \ref{2-10}. The power spectrum of Figure \ref{2-10}(a) is shown in Figure \ref{2-10}(b) with the matrix and precipitate diffraction spots and the measured d-values and angles of the precipitate are tabulated in Table \ref{tab:2-10} (d\textsubscript{1}=3.00, d\textsubscript{2}=1.85, d\textsubscript{3}=2.61, $\alpha$\textsubscript{12}=35$^\circ$, $\alpha$\textsubscript{13}=55$^\circ$, $\alpha$\textsubscript{23}=90$^\circ$). The JEMS simulations(Figure \ref{2-10}(c)) show that $<2\bar10>$ and $<1\bar20>$ (Miller notation) family of directions of Y\textsubscript{4}Zr\textsubscript{3}O\textsubscript{12} can produce the diffraction pattern with the same d-spacings and angular relationships as in Figure \ref{2-10}(b). In Miller - Bravais notation, this is equivalent to $<5\bar4\bar10>$ and $<4\bar510>$ directions. (For convenience to compare with existing literature, we are following the miller notation throughout this study. Subscripts '$\delta$' and '$\alpha$' are used to differentiate planes and directions corresponding to Y\textsubscript{4}Zr\textsubscript{3}O\textsubscript{12} and bcc Fe matrix respectively.)
\par The individual directions in a family are the same in all respects, related to each other by symmetry of Y\textsubscript{4}Zr\textsubscript{3}O\textsubscript{12} unitcell and hence can be generated using the six space group generators ($1; 3^+ 0,0,z;$ $3^- 0,0,z;$ $ -1 0,0,0;$ $-3^+ 0,0,z;$ $-3^- 0,0,z $) of $R\bar3$ space group. The $[1\bar20]_\delta$, $[230]_\delta$, $[3\bar10]_\delta$, $[\bar120]_\delta$, $[\bar2\bar30]_\delta$ and $[310]_\delta$ directions belong to $<1\bar20>_\delta$ family and $[130]_\delta$, $[\bar1\bar30]_\delta$, $[2\bar10]_\delta$, $[\bar210]_\delta$, $[320]_\delta$ and $[\bar3\bar20]_\delta$  belong to the $<2\bar10>_\delta$ family. The diffraction spots of Y\textsubscript{4}Zr\textsubscript{3}O\textsubscript{12} shown in Figure \ref{2-10}(b) is widely observed in ODS steels containing Zr and is indexed as $[2\bar10]_\delta$ zone axis by various authors\cite{dong2017enhancement,xu2017microstructure,dou2020effects,dou2014tem,dou2019morphology}.

\par
\begin{figure}[!hbt]
	\centering
	\subfigure[]{\includegraphics[width=0.45\textwidth]{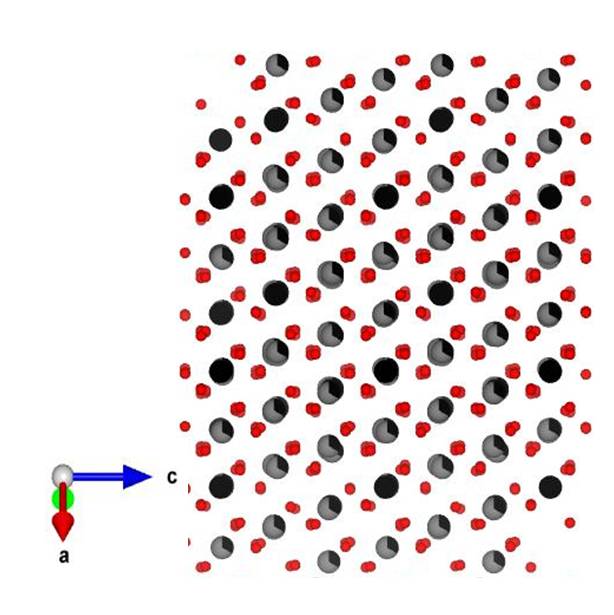}}
	\subfigure[]{\includegraphics[width=0.45\textwidth]{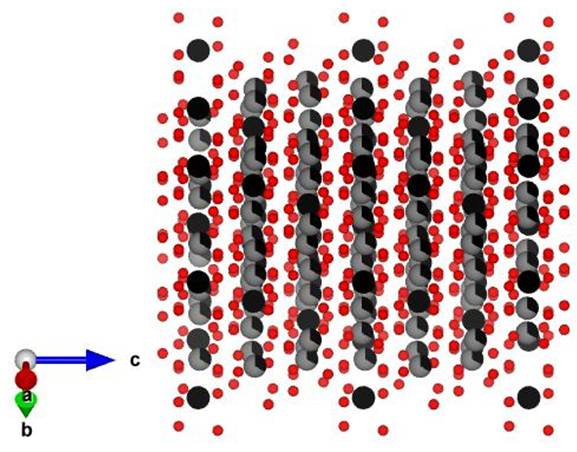}}
	\caption{Atomic arrangement along (a)$<1\bar20>$ and (b)$<2\bar10>$ directions of Y\textsubscript{4}Zr\textsubscript{3}O\textsubscript{12}. Black, grey and red spheres represent Zr, Y and O atoms respectively. }
	\label{2-10_fig}
\end{figure}

 However, a close examination of these directions shows that even though the d-spacings of diffracting planes can be the same, these families of directions are different in terms of atomic arrangement and close packing. Figures \ref{2-10_fig} (a) and (b) show the atomic arrangement of $<1\bar20>_\delta$ and $<2\bar10>_\delta$ respectively. $<1\bar20>_\delta$ is closely packed than $<2\bar10>$. One of the major requirements for an OR to be energetically favourable and to occur frequently is that the close packed directions of the structures involved should align in the same direction\cite{dahmen1982orientation}. Therefore, it is reasonable to index the diffraction spots for Y\textsubscript{4}Zr\textsubscript{3}O\textsubscript{12} in the power spectrum, Figure \ref{2-10}(b) and simulated diffraction pattern Figure \ref{2-10}(c) as close packed $<1\bar20>_\delta$. 
Another way to differentiate the contrast from both the crystal directions is by multislice simulations.
\section{Multislice simulations along $<1\bar20>_\delta$ and $<2\bar10>_\delta$ zone axes}
High resolution electron micrographs are interference patterns arising from e{\tiny }lectron-specimen interactions which are further modified by instrumental aberrations and are not a direct mapping of the specimen lattice. Therefore, direct interpretation of contrasts in the experimental data can be erroneous and  they have to be complemented by numerical simulations, like multislice simulations. 
\par
\begin{figure}[!hbt]
	\centering
	\subfigure[]{\includegraphics[width=0.48\textwidth]{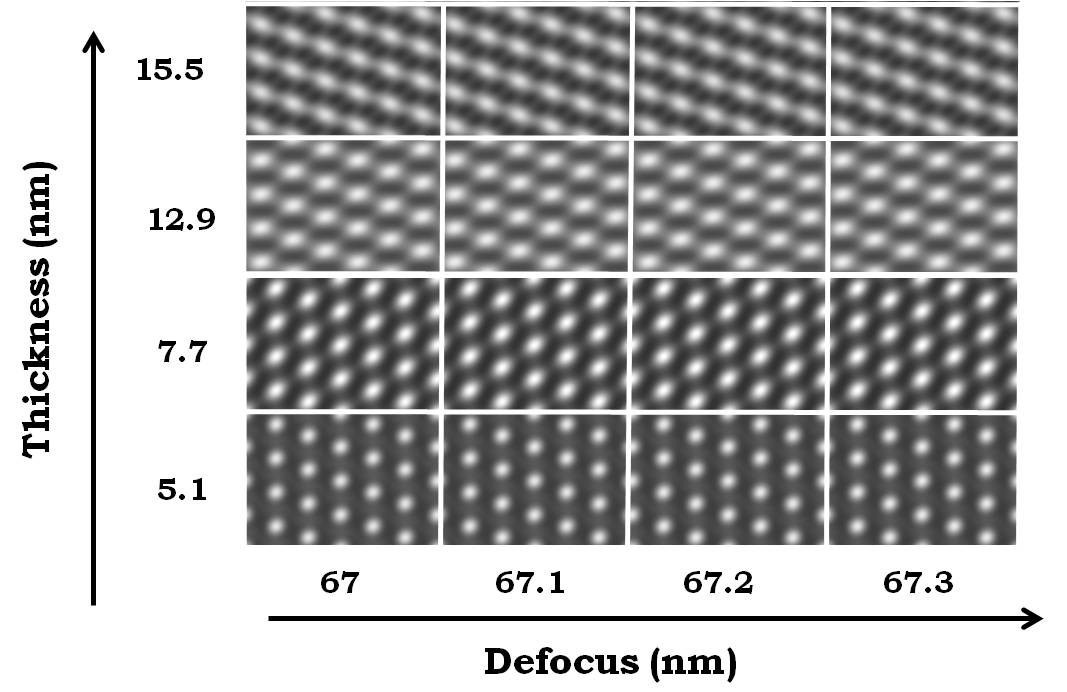}}
	\subfigure[]{\includegraphics[width=0.48\textwidth]{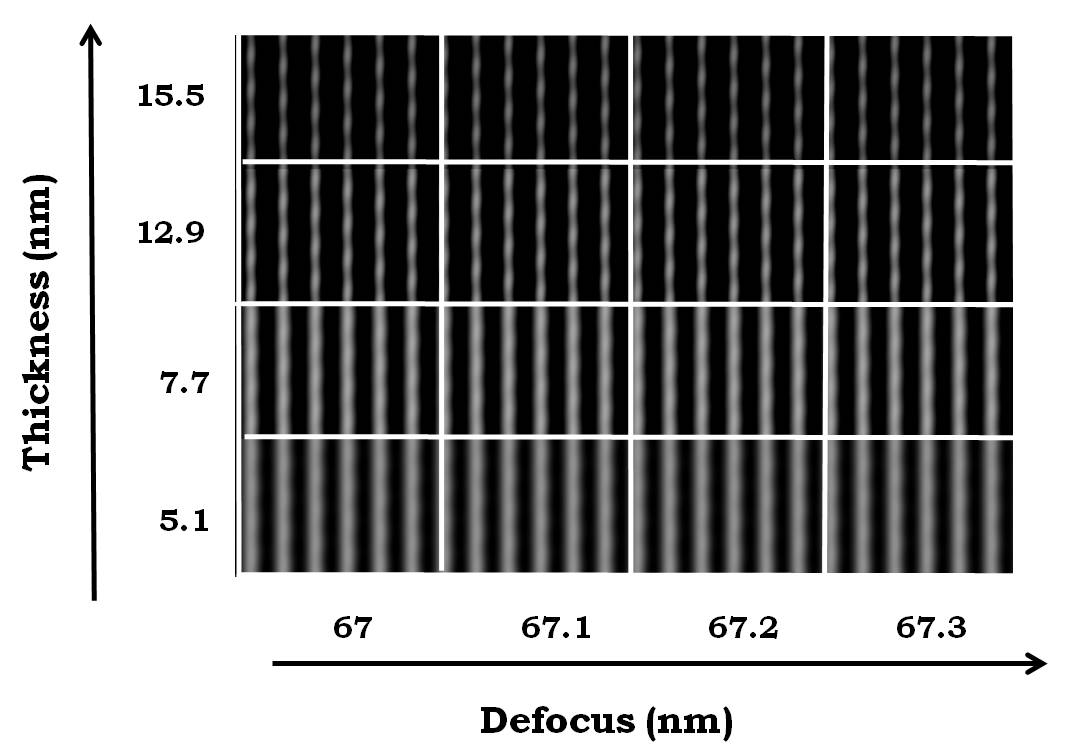}}	
	\caption{Defocus thickness map for (a) $<1\bar20>$ and (b) $<2\bar10>$ zone axes of Y\textsubscript{4}Zr\textsubscript{3}O\textsubscript{12}(C\textsubscript{s}, C\textsubscript{c}=1.2 nm)}
	\label{defocus}
\end{figure}

 To further understand the contrasts of $<1\bar20>_\delta$ and $<2\bar10>_\delta$, multislice simulations were carried out using JEMS software for various thicknesses and defocus values in these directions. The experimental images were taken at defocus of $\sim$ 67 nm with C\textsubscript{s}=1.2 nm and C\textsubscript{c}=1.2 nm. The defocus thickness map for $<1\bar20>$ and $<2\bar10>$ of   Y\textsubscript{4}Zr\textsubscript{3}O\textsubscript{12} shown in Figure \ref{defocus}(a) and (b) are entirely different for all the thickness and defocus ranges relevant in the current experiment. The $<1\bar20>_\delta$ directions give a `dotted contrast' while the $<2\bar10>_\delta$ give rise to a `line contrast'. For $<1\bar20>$ directions, though the variation in contrast with respect to defocus is negligible while the contrast varies considerably with respect to thickness. 
However, it is obvious that the contrasts observed in precipitates of Figure \ref{cluster} and Figure \ref{2-10} are in exact agreement with the simulated HR pattern of $<1\bar20>$ of Y\textsubscript{4}Zr\textsubscript{3}O\textsubscript{12} of the thickness range, 5-12 nm, which confirms that the observed direction is $<1\bar20>_\delta$.

\FloatBarrier
%
%
\par Therefore, the predominant precipitate matrix OR deduced from Figures \ref{cluster} and \ref{2-10} is:
\begin{equation}
[1\bar20]_\delta \big|\big|[111]_\alpha\text{  and  }  (21\bar2)_\delta \big|\big|(\bar110)_{\alpha}
\label{OR1}
\end{equation}
In this OR, the $(21\bar2)$ plane (d\textsubscript{$21\bar2$}=2.61 {\AA}) of precipitate is oriented parallel to $(\bar110)$ plane (d\textsubscript{$\bar110$}=2.02 {\AA}) of $\alpha$-Fe. The angle between these two planes is $\sim0.3^\circ$. 
The (003) plane with d value d\textsubscript{$003$}=3.03 {\AA} is nearly parallel to $(01\bar1)$ plane of the matrix, with an angle $\sim$ 5$^\circ$. The angle between $(211)$ plane of the precipitate and the $(1\bar10)$ plane of the matrix is also $\sim$ 5 $^\circ$.

All the precipitates in Figure \ref{cluster} and 86\% of the precipitates analyzed in this study possess this orientation relationship. The existence of a predominant precipitate-matrix relationship points to the fact that the oxide powders completely dissolves in the matrix during the ball milling process and nucleates as a new phase in a preferred direction of the matrix. 

\par The alternate ways to write this orientation relationship are :
\begin{align*}
&[310]_\delta||[111]_\alpha \text{  and  } (1\bar3\bar2)_\delta||(\bar110)_\alpha;\\
&[\bar3\bar10]_\delta||[111]_\alpha \text{  and  } (\bar132)_\delta||(\bar110)_\alpha;
\end{align*}
\begin{align*}
&[1\bar20]_\delta||[111]_\alpha \text{  and  } (2\bar12)_\delta||(\bar110)_\alpha;\\
&[\bar120]_\delta||[111]_\alpha \text{  and  } (21\bar2)_\delta||(\bar110)_\alpha;
\end{align*}
\begin{align*}
&[\bar2\bar30]_\delta||[111]_\alpha \text{  and  } (3\bar22)_\delta||(\bar110)_\alpha;\\
&[230]_\delta||[111]_\alpha \text{  and  } (\bar32\bar2)_\delta||(\bar110)_\alpha
\end{align*}
Since the bcc structure has six \{110\} planes and each of those contains two [111] directions, twelve equivalent combinations of OR1 are probable, each one is referred to as a `variant'.

\FloatBarrier

\section{Misfit dislocations and strain contrast at the precipitate matrix interface}

\par The relationship between the planes of the Y\textsubscript{4}Zr\textsubscript{3}O\textsubscript{12} and Fe matrix in orientation relationship described by equation \ref{OR1} is: $4d_{\bar110,\alpha}\simeq 3d_{21\bar2,\delta}$,
which makes it semi-coherent with the matrix. The lattice mismatch, $\Delta =\Big| \frac{d_{\delta, 21\bar2}-d_{\alpha,\bar110}}{d_{\delta, 21\bar2}}\Big| =29.8\% $ and the bright shadow extending to $\sim$ 1 nm surrounding the precipitate is the reflection of the coherency strain arising due to this lattice mismatch. The misfit strain between the precipitate and matrix is usually accommodated by misfit dislocations in between the coherent interface patches, until a critical size of precipitate is reached. 
\par
\begin{figure}[!hbt]
	\centering
	\includegraphics[width=\linewidth]{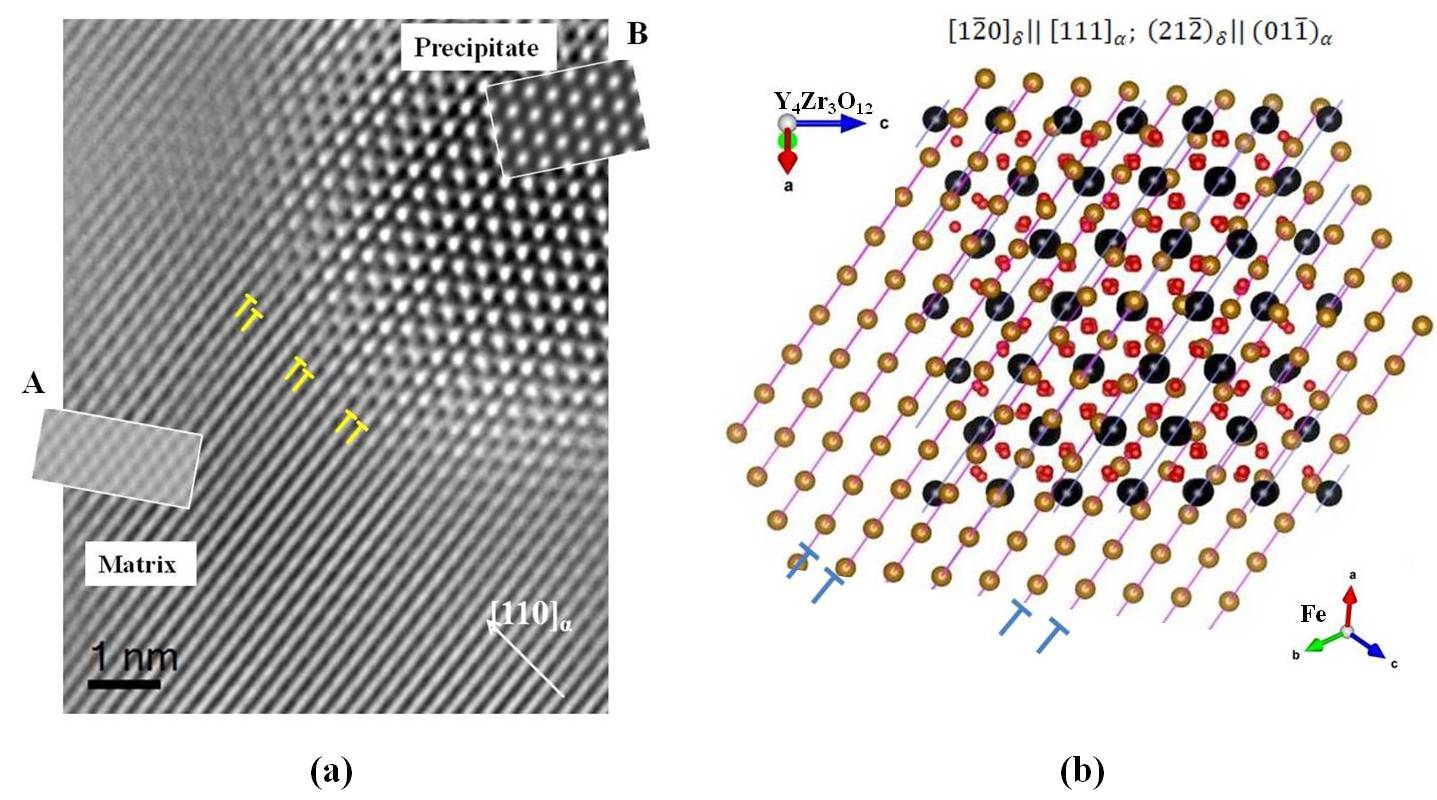}
	\caption{The misfit dislocation substructure of OR1-interface. (a) The precipitate matrix interface of precipitate in Figure \ref{2-10}. The simulated HREM map of precipitate and matrix is shown in regions marked A and B respectively (thickness$\sim$ 7 nm, C\textsubscript{c}, C\textsubscript{s}=1.2 nm, defocus=67 nm). the precipitate matrix interface has periodic array of dislocations, marked in yellow.\\
	(b) Atomic model of OR $[1\bar20]_\delta||[\bar111]_\alpha$ and $(21\bar2)_\delta||(\bar110)_\alpha$ projected along $[1\bar20]_\delta$/$[111]_\alpha$. Black, red and brown spheres are Y/Zr, O and Fe atoms respectively. The $(21\bar2)_\delta$ and $(\bar110)_\alpha$ planes are marked by blue lines and pink lines respectively. The misfit dislocations arising when one set of $(\bar110)_\alpha$ matches exactly with $(21\bar2)_\delta$ are marked in blue. The possible dislocation arrays are in exact agreement with experimental observation, (a).}
	\label{interface}
\end{figure}

 The noise filtered image of the precipitate matrix interface of Figure \ref{2-10}(a) is shown in Figure \ref{interface}. The simulated HREM map for precipitate and matrix is in agreement with the observed HRTEM image and in the case of precipitates, the interference contrast is arising from cation ordering.  The simulated patterns of precipitate and the matrix are shown in regions marked A and B respectively. A periodic array of geometric dislocations can be seen at the precipitate - matrix interface, which is marked in yellow. Some precipitate matrix planes are perfectly coinciding. After a set of three such coinciding planes, two matrix ($\bar110$) planes terminate on both sides of a $(21\bar2)$ precipitate plane, creating a dislocation doublet. Each dislocation doublet is separated by a distance of $\sim$ 0.9 nm. 

The strain due to lattice mismatch is accommodated by these misfit dislocation doublets. 

\par The periodic strained and un-strained regions in Figure \ref{interface} gives rise to strain-contrast as ripples along the precipitate direction parallel to $[\bar110]_\alpha$ (visible clearly in Figure \ref{2-10}(a)). The periodicity of these ripples are same as the periodicity of the misfit Moire fringes\cite{ribis2013influence}, given by:$\Big|\frac{d_1d_2}{d_1-d_2}\Big|=0.87$ nm. Since this spacing is same as the spacing between dislocation doublets, the misfit Moire fringes have been previously used to predict the spacing between misfit dislocations\cite{carter2016transmission,ribis2012interfacial}. 
\par Further, an atomic model is constructed for the orientation relationship $[1\bar20]_\delta||[111]_\alpha$ and $(\bar110)_\alpha||(21\bar2)_\alpha$ and is shown in Figure \ref{interface}(b), projected along $[1\bar20]_\delta$/$[111]_\alpha$ direction. The Y/Zr, O and Fe atoms are represented by  black, red and brown spheres and the $(21\bar2)_\delta$ and $(\bar110)_\alpha$ planes are marked by blue lines and pink lines respectively. The possible array of misfit dislocations in $(\bar110)_\alpha$/$(21\bar2)_\delta$ boundary (marked in blue) are in exact agreement with experimental observation shown in Figure \ref{interface}.  
The strain contrast ripples are along highly packed precipitate direction parallel to $[\bar110]_\alpha$ and it is similar to the orthogonal strains along $<001>_\alpha$ and $<110>_\alpha$ directions for predominant bcc/hcp orientation relationships\cite{dahmen1982orientation}.

\FloatBarrier
\section{Orientation related by rotation to preferred orientation}
\begin{figure}[!hbt]
	\centering
	\includegraphics[width=\textwidth]{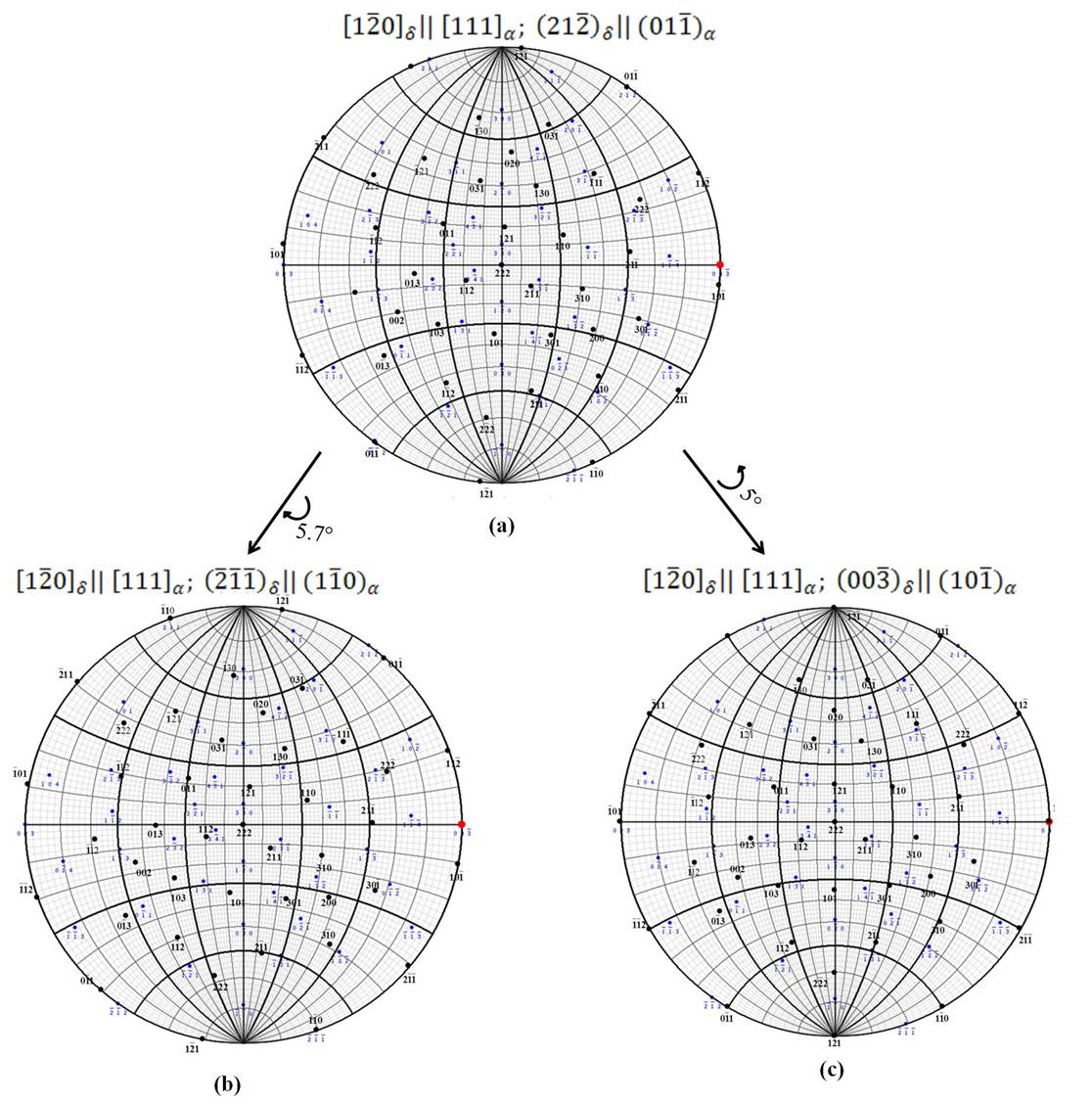}
	\caption{Combined stereographic projection of $[1\bar20]_\delta$ and $[111]_\alpha$ with $(21\bar2)_\delta || (\bar110)_\alpha$. Blue spots indicate $[1\bar20]$ zone axis of Y\textsubscript{4}Zr\textsubscript{3}O\textsubscript{12} and black spots correspond to [111] zone axis of Fe}
	\label{sterogram}
\end{figure}
A combined stereographic projection of $[1\bar20]_\delta$ and $[111]_\alpha$ with $(21\bar1)_\delta || (\bar110)_\alpha$ is given in Figure \ref{sterogram}(a) in which, it is evident that the angle between $(003)_\delta$ and $(\bar101)_\alpha$ is $\sim$ 5$^\circ$. A relative rotation by 5$^\circ$ with  $[111]_\alpha||[1\bar20]_\delta$ condition intact will bring $(00\bar3)_\delta$ and $(\bar101)_\alpha$ in coincidence (Figure \ref{sterogram}(c)). A few precipitates with this new orientation, with the same directional relationship as the most preferred orientation but different parallel planes, are also observed in Zr-ODS. 
\par
\begin{figure}[!hbt]
	\centering
	\subfigure[]{\includegraphics[width=0.32\linewidth]{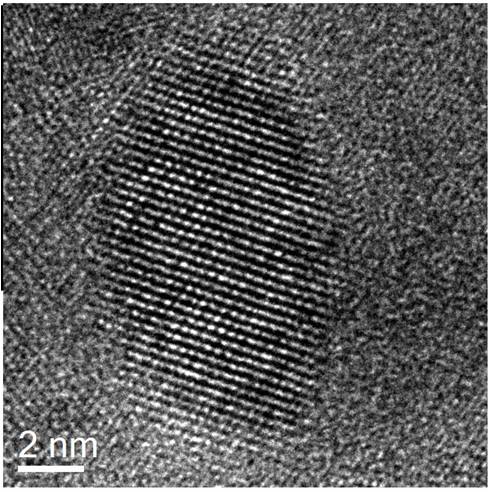}}
	\subfigure[]{\includegraphics[width=0.32\linewidth]{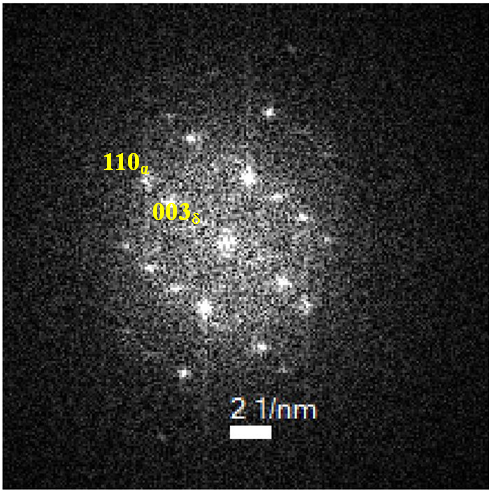}}
	\subfigure[]{\includegraphics[width=0.32\linewidth]{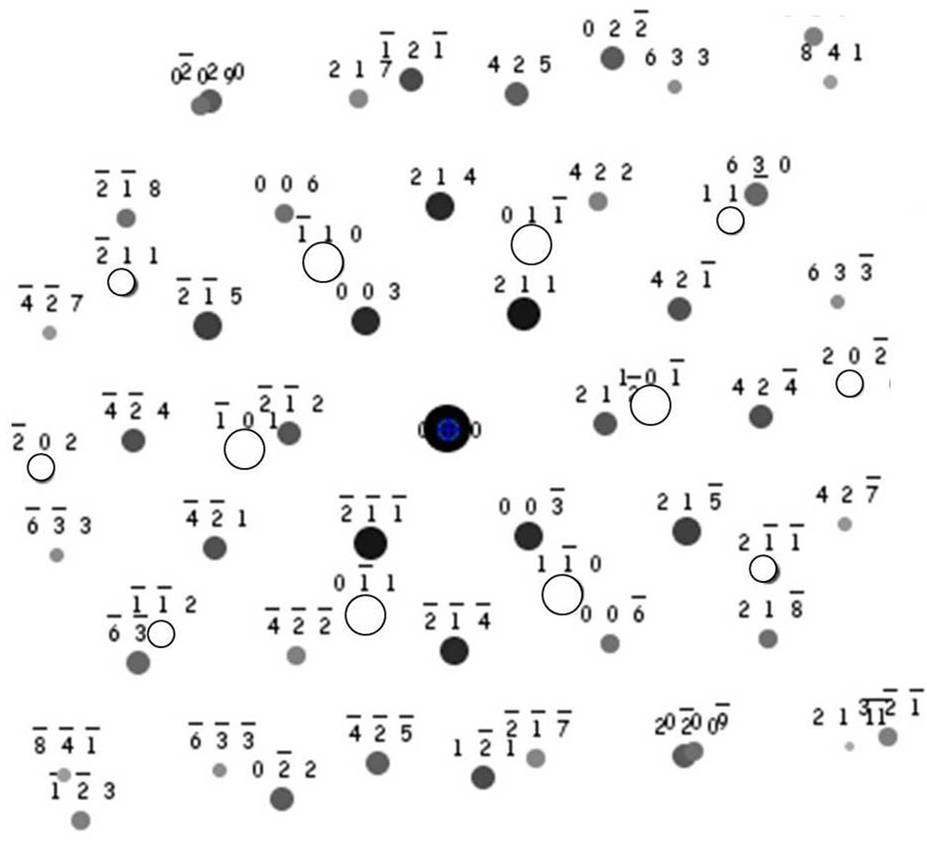}}
	\captionof{figure}{(a) HRTEM micrograph of a Y\textsubscript{4}Zr\textsubscript{3}O\textsubscript{12} precipitate (b) The power spectrum of (a). (c) A combined diffraction pattern of $[111]_{Fe}||[1\bar20]_\delta$ with $(110)_{Fe}||(00\bar3)_{\delta}$. The matrix spots are shown as open circles.}
	\label{OR_2}
	\bigskip
	\captionof{table}{Inter-planar distances (d) and angles ($\alpha$) of the nanoparticle of Figure \ref{OR_2}}
	\bigskip
	\begin{tabular}{cccccccc}
		\toprule
		\medskip
		&d(\AA),$\alpha(^\circ)$ &$d_1$ &$d_2$ &$d_3$ &$\alpha_{12}$ &$\alpha_{13}$ &$\alpha_{23}$\\\midrule
		\medskip
		&Measured &3.00  &1.85   &2.61  &35.2  &54.6 &90\\
		\medskip
		Particle &Planes	&$\left(00\bar3\right)$	&$\left(12\bar4\right)$		&$\left(\bar1\bar2\bar2\right)$		&-	&-	&- \\ 
		\medskip
		&Calculated    &3.03	&1.85 	&2.61 	&35 	&55	&90.59 \\\midrule
		\medskip
		&Measured &2.04  &2.04 &2.04  &60  &120  &60\\
		\medskip
		Matrix  &Planes	&$\left(01\bar1\right)$	&$\left(10\bar1\right)$	&$\left(1\bar10\right)$	&-	&-	&- \\ 
		\medskip
		&Calculated    &2.02 &2.02	&2.02 	&60 	&120	&60\\\bottomrule
		\medskip
	\end{tabular}
	\label{tab:OR_2}
\end{figure}

 A Y\textsubscript{4}Zr\textsubscript{3}O\textsubscript{12} precipitate with $[1\bar20]_\delta||[111]_\alpha$ and $(003)_\delta||(\bar101)_\alpha$ is shown in Figure \ref{OR_2}(a). The power spectrum of the precipitate and the matrix is shown in Figure \ref{OR_2}(b). The combined diffraction pattern of $[1\bar20]_\delta$ and $[111]_\alpha$ is shown in Figure \ref{OR_2}(b). The mismatch with matrix d-value ($d_{\bar101}=2.02\; {\AA}$) and the precipitate d-value ($d_{300}=3.03$) is, $\Delta =\Big| \frac{d_{\delta, 00\bar3}-d_{\alpha,\bar110}}{d_{\delta, 00\bar3}}\Big| =33\% $ and $\Big|{2d_{\delta, 00\bar3}-3d_{\alpha,\bar110}}\Big|=0$, which ensures a perfect match of every third matrix plane to the second precipitate plane.

The orientation relationships equivalent to $[1\bar20]_\delta||[111]_\alpha$ and $(00\bar3)_\delta||(\bar101)_\alpha$ are:
\begin{align*}
&[310]_\delta||[111]_\alpha \text{  and  } (00\bar3)_\delta||(\bar101)_\alpha;\\
&[\bar3\bar10]_\delta||[111]_\alpha \text{  and  } (00\bar3)_\delta||(\bar101)_\alpha;\\
\end{align*}
\begin{align*}
&[1\bar20]_\delta||[111]_\alpha \text{  and  } (00\bar3)_\delta||(\bar101)_\alpha;\\
&[\bar120]_\delta||[111]_\alpha \text{  and  } (00\bar3)_\delta||(\bar101)_\alpha;
\end{align*}
\begin{align*}
&[\bar2\bar30]_\delta||[111]_\alpha \text{  and  } (00\bar3)_\delta||(\bar101)_\alpha;\\
&[230]_\delta||[111]_\alpha \text{  and  } (00\bar3)_\delta||(\bar101)_\alpha\\
\end{align*}
With six {110}$_\alpha$ planes each containing two $[111]_\alpha$ directions, twelve variants of OR2 are possible. 
\par
\begin{figure}[!hbt]
	\centering
	\includegraphics[width=\linewidth]{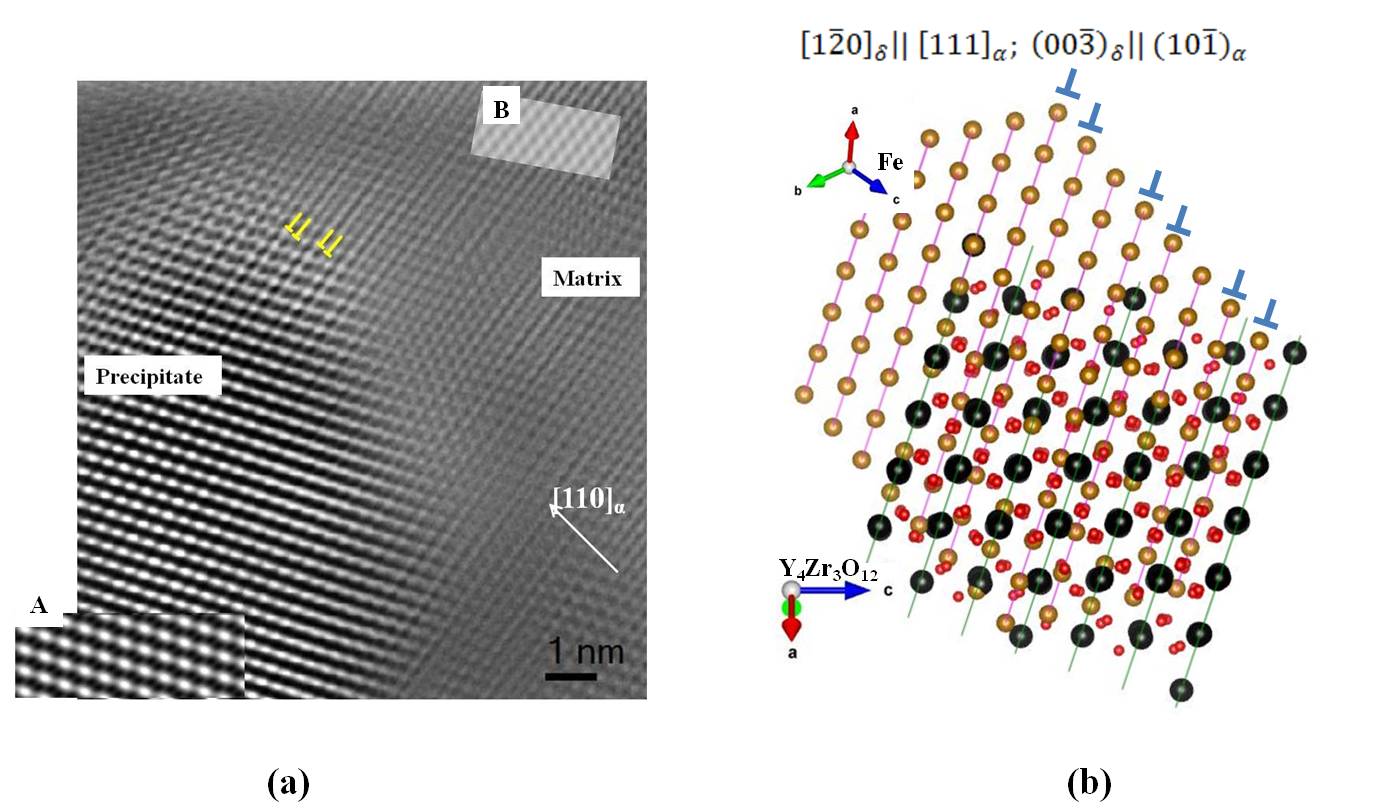}
	\caption{(a) The precipitate matrix interface structure of precipitate in Figure \ref{OR_2}. The simulated diffraction patterns of precipitate and matrix are given in A and B respectively (defocus=67 nm, thickness $\sim$ 15 nm, C\textsubscript{c}, C\textsubscript{s}=1.2 nm).\\
	(b) Atomic model of OR $[1\bar20]_\delta||[\bar111]_\alpha$ and $(003)_\delta||(\bar110)_\alpha$ projected along $[1\bar20]_\delta$. Black, red and brown spheres are Y/Zr, O and Fe atoms respectively. The $(003)_\delta$ and $(\bar110)_\alpha$ planes are marked by green and pink lines respectively. The misfit dislocations arising when one set of $(\bar110)_\alpha$ matches exactly with $(003)_\delta$ are marked in blue. The possible dislocation arrays are in exact agreement with (a)}
	\label{interface_2}
\end{figure}

 The periodicity of the dislocations in the interface of precipitate with this OR is depicted in noise filtered image Figure \ref{interface_2}. The simulated HR map for $[1\bar20]_\delta$ and $[111]_\alpha$ are superimposed in regions marked A and B respectively . The spacing between dislocation doublets is 0.6 nm, which is same as the misfit Moire fringe spacing; $\Big|\frac{d_1d_2}{d_1-d_2}\Big|=0.61$ nm.

An atomic model of OR $[1\bar20]_\delta||[\bar111]_\alpha$ and $(003)_\delta||(\bar110)_\alpha$ along $[1\bar20]_\delta$ is shown in Figure \ref{interface_2}(b). Black, red and brown spheres are Y/Zr, O and Fe atoms respectively. The $(003)_\delta$ and $(\bar110)_\alpha$ planes are marked by green and pink lines respectively. The misfit dislocations arising when one set of $(\bar110)_\alpha$ matches exactly with $(003)_\delta$ are marked in blue. The possible dislocation arrays are in exact agreement with the experimental observation shown in Figure \ref{interface_2}.

\section{Comparison of observed ORs with reported ORs in hcp/bcc system}
The most commonly reported ORs in hcp/bcc precipitate-matrix interfaces are: Burger's OR \cite{burgers1934process}, Pitsche-Schrader (PS) OR\cite{rong1984crystallography} and Potter's OR\cite{rong1984crystallography,potter1973structure}. All of these form low energy, semi-coherent interfaces. The Burger's, Potters and Pitsche-Schrader ORs are interrelated and can be generated from each other using small relative rotations\cite{matsukawa2019crystallography}. A relative rotation of the coinciding direction of Burger's OR by an angle of 5.26$^\circ$ gives the PS OR and a A rotation of Burger's OR by $\sim$ 2$^\circ$ results in Potter's OR.

\par Similarly, in our trigonal/bcc system, a rotation of the combined stereographic projection, Figure \ref{sterogram}(a) in clockwise direction by 5.7 $^\circ$ makes $(\bar110)_\alpha$ and $(211)_\delta$ (d=3.00 {\AA}) parallel to each other (see Figure \ref{sterogram}(b)). Being close packed planes in the respective crystal systems, this might be an energetically possible OR. 
Due to the closeness of d value of $(211)_\delta$ and $(00\bar3)_\delta$ planes, the interface structure will be similar to that of $[1\bar20]_\delta||[\bar111]_\alpha$ and $(003)_\delta||(\bar110)_\alpha$, with a dislocation doublet spacing of 0.6 nm. This OR can also occur in twelve variants and each variant can be expressed in six different ways, given by:
\begin{align*}
[310]_\delta||[111]_\alpha \text{  and  } (1\bar31)_\delta||(\bar110)_\alpha;\\
[\bar3\bar10]_\delta||[111]_\alpha \text{  and  } (1\bar31)_\delta||(\bar110)_\alpha;\\
\end{align*}
\begin{align*}
[1\bar20]_\delta||[111]_\alpha \text{  and  } (211)_\delta||(\bar110)_\alpha;\\
[\bar120]_\delta||[111]_\alpha \text{  and  } (\bar2\bar1\bar1)_\delta||(\bar110)_\alpha;
\end{align*}
\begin{align*}
[\bar2\bar30]_\delta||[111]_\alpha \text{  and  } (3\bar2\bar1)_\delta||(\bar110)_\alpha;\\
[230]_\delta||[111]_\alpha \text{  and  } (\bar321)_\delta||(\bar110)_\alpha\\
\end{align*}

\FloatBarrier
\section{Random orientations of precipitates in the matrix}
In addition to the ORs mentioned in previous sections, two random precipitate-matrix orientations with high order zone axes of Y\textsubscript{4}Zr\textsubscript{3}O\textsubscript{12} aligned to [111] direction of bcc are also observed in Zr-ODS. These orientations are seen in relatively large precipitates, possess a relationship with matrix-plane and can be expressed in the form of complete orientation relationships. 

\par The HRTEM image of one such precipitate in Zr-ODS is shown in Figure \ref{still022}. The measured and calculated d- spacings of both the particles and matrix are tabulated in Table \ref{tab:still022}. The precipitate is Y\textsubscript{4}Zr\textsubscript{3}O\textsubscript{12} with zone axis $[\bar4$7 10]. The $(\bar110)$ plane of the matrix is nearly parallel to $(12\bar1)$ plane of Y\textsubscript{4}Zr\textsubscript{3}O\textsubscript{12}. The d-values of these planes are 2.01 {\AA} and 3.00 {\AA} respectively. Due to the larger size of the precipitates,  the elastic strain is higher unlike the small precipitates of the previous cases. The strain cannot be accommodated by misfit dislocations alone and therefore, the interface structure appears complicated. Extensive twinning of the precipitate planes can be observed in the precipitate and  interface. 
 
This OR is observed only in 2\% of the precipitate analyzed. The angle between close packed direction, $[1\bar20]$ and $[\bar47\;10]$ is $\sim$ 56$^\circ$. 
Due to the symmetry of matrix and precipitates, the power spectrum in Figure \ref{still022} (b) is similar to two families of directions in Y\textsubscript{4}Zr\textsubscript{3}O\textsubscript{12}, $<\bar47\;10>_\delta$ and $<\bar74\;10>_\delta$ and hence this orientation relationship can be written in 12 different ways.

\begin{align*}
&[\bar47\;\;10]_\delta||[111]_\alpha \text{  and  } (12\bar1)_\delta||(\bar110)_\alpha;\\
&[4\bar7\;\;\bar{10}]_\delta||[111]_\alpha \text{  and  } (12\bar1)_\delta||(\bar110)_\alpha;\\
&[7\bar4\;\;\bar{10}]_\delta||[111]_\alpha \text{  and  } (211)_\delta||(\bar110)_\alpha;\\
&[\bar74\;\;10]_\delta||[111]_\alpha \text{  and  } (211)_\delta||(\bar110)_\alpha;
\end{align*}
\begin{align*}
&[11\;4\;10]_\delta||[111]_\alpha \text{  and  } (\bar231)_\delta||(\bar110)_\alpha;\\
& [\bar{11}\;\bar4\;\bar{10}]_\delta||[111]_\alpha \text{  and  } (\bar231)_\delta||(\bar110)_\alpha; \\
&[\bar4\;\bar{11}\;10]_\delta||[111]_\alpha \text{  and  } (\bar321)_\delta||(\bar110)_\alpha;\\
&[4\;11\;\bar{10}]_\delta||[111]_\alpha \text{  and  } (\bar321)_\delta||(\bar110)_\alpha
\end{align*}
\begin{align*}
&[\bar7\;\bar{11}\;10]_\delta||[111]_\alpha \text{  and  } (12\bar1)_\delta||(\bar110)_\alpha;\\
&[7\;11\;\bar{10}]_\delta||[111]_\alpha \text{  and  } (12\bar1)_\delta||(\bar110)_\alpha;\\
&[\bar7\;\bar{11}\;\;\bar{10}]_\delta||[111]_\alpha \text{  and  } (211)_\delta||(\bar110)_\alpha;\\
&[11\;7\;10]_\delta||[111]_\alpha \text{  and  } (211)_\delta||(\bar110)_\alpha;
\end{align*}

\begin{figure}[!hbt]
	\centering
	\subfigure[]{\includegraphics[width=0.36\linewidth]{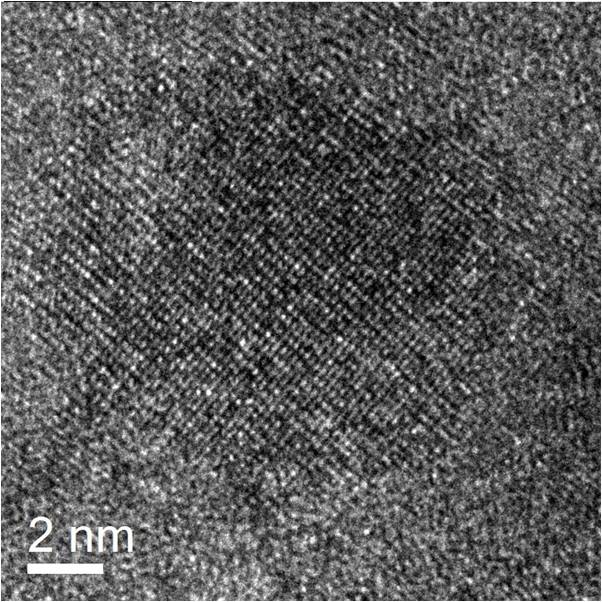}}
	\subfigure[]{\includegraphics[width=0.36\linewidth]{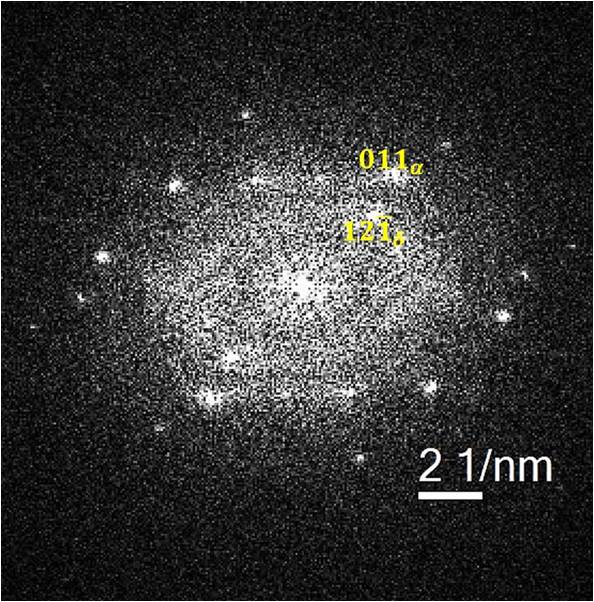}}
	\subfigure[]{\includegraphics[width=0.36\linewidth]{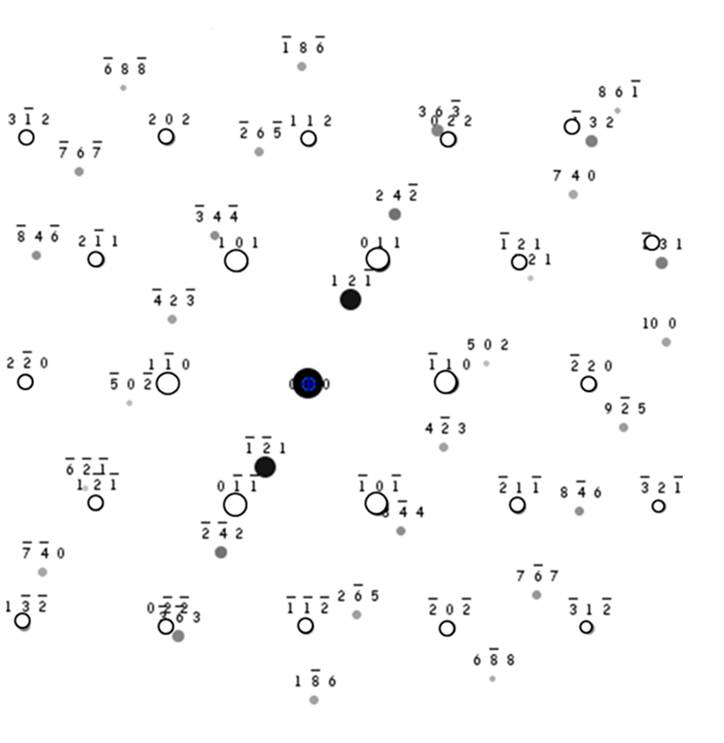}}
	\subfigure[]{\includegraphics[width=0.36\linewidth]{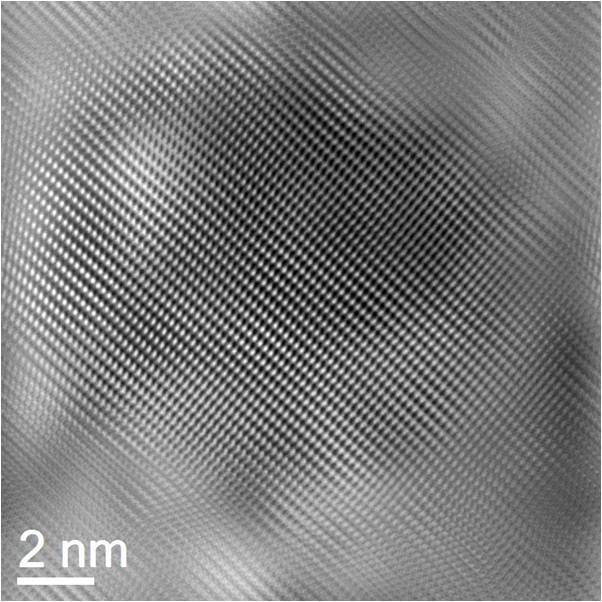}}
	\captionof{figure}{(a) HRTEM micrograph of a Y\textsubscript{4}Zr\textsubscript{3}O\textsubscript{12} in Zr-ODS. (b) Power spectrum of the micrograph with corresponding (hkl) values. The zone axis is $[\bar47 \;\;10]$ (c) Simulated diffraction pattern of the $[\bar47\;\; 10]$ zone axis of Y\textsubscript{4}Zr\textsubscript{3}O\textsubscript{12} with [111] zone axis of $\alpha$-Fe. The open circles corresponds to matrix spots and the filled circles corresponds to precipitate-spots. (d) IFFT constructed from power spectrum using precipitate spots.}
	\label{still022}
	\captionof{table}{Inter-planar distances (d) and angles ($\alpha$) of the nanoparticle of Figure \ref{still022}}
	\bigskip
	\begin{tabular}{cccccccc}
		\toprule
		\medskip
		&d(\AA),$\alpha(^\circ)$ &$d_1$ &$d_2$ &$d_3$ &$\alpha_{12}$ &$\alpha_{13}$ &$\alpha_{23}$\\\midrule
		\medskip
		&Measured &1.56  &3.0   &1.91  &56.0  &31.2 &88\\
		\medskip
		&Planes	&$\left(\bar50\bar2\right)$	&$\left(\bar1\bar21\right)$		&$\left(\bar42\bar3\right)$		&-	&-	&- \\ 
		\medskip
		&Calculated    &1.58	&3.01 	&1.90 	&56.3 &31.7	&88.01 \\\bottomrule
		\medskip
	\end{tabular}
	\label{tab:still022}
\end{figure}

\par The Figure \ref{still024}(a) shows a spherical Y\textsubscript{4}Zr\textsubscript{3}O\textsubscript{12} precipitate of diameter $\sim$ 12 nm. Careful analysis of power spectrum reveals that 
the spots marked in \ref{still024} (b) are in agreement with the simulated pattern displayed in \ref{still022}(c). The zone axis is $[\bar4$1  $\bar{11}]$. The $(31\bar1)$ plane of the precipitate (d=2.26) is parallel to $(\bar110)$ plane of the matrix (d=2.04). The lattice mismatch is $\sim$ 9.7\%. In this orientation, the equivalent ORs are:
\begin{align*}
&[\bar41\;\bar{11}]_\delta||[111]_\alpha \text{  and  } (31\bar1)_\delta||(\bar110)_\alpha;\\
&[4\bar1\;11]_\delta||[111]_\alpha \text{  and  } (31\bar1)_\delta||(\bar110)_\alpha;\\
&[1\bar4\;11]_\delta||[111]_\alpha \text{  and  } (131)_\delta||(\bar110)_\alpha;\\
&[\bar14\;\;11]_\delta||[111]_\alpha \text{  and  } (131)_\delta||(\bar110)_\alpha;
\end{align*}
\begin{align*}
&[1\;5\;11]_\delta||[111]_\alpha \text{  and  } (4\bar31)_\delta||(\bar110)_\alpha;\\
& [\bar1\;\bar5\;\bar{11}]_\delta||[111]_\alpha \text{  and  } (4\bar31)_\delta||(\bar110)_\alpha; \\
&[\bar5\;\bar1\;11]_\delta||[111]_\alpha \text{  and  } (3\bar41)_\delta||(\bar110)_\alpha;\\
&[5\;1\;\bar{11}]_\delta||[111]_\alpha \text{  and  } (3\bar41)_\delta||(\bar110)_\alpha
\end{align*}
\begin{align*}
&[5\;4\;\bar{11}]_\delta||[111]_\alpha \text{  and  } (\bar141)_\delta||(\bar110)_\alpha;\\
&[\bar{5}\;\bar4\;11]_\delta||[111]_\alpha \text{  and  } (\bar141)_\delta||(\bar110)_\alpha;\\
&[4\;5\;11]_\delta||[111]_\alpha \text{  and  } (\bar411)_\delta||(\bar110)_\alpha;\\
&[\bar4\;\bar5;\bar{11}]_\delta||[111]_\alpha \text{  and  } (\bar411)_\delta||(\bar110)_\alpha;
\end{align*}
\par The orientation relationships and their nature of interface and frequency of occurrence are tabulated in Table \ref{tab:ORs}.

\begin{table}[!hbt]
	\ra{1.3}
	\captionof{table}{Summary of the orientation relationships found in as prepared Zr-ODS steel.}
	\begin{center}
		\begin{tabular}{cp{3cm}p{3cm}p{3cm}p{2cm}}
			\toprule
			Sl.No.&Orientation Relationships & Nature of interface &Relationship between planes&Frequency of occurrence(\%)\\\midrule
			1&$[1\bar20]_{\delta}\;\big|\big|\;[111]_\alpha$ & semi-coherent & $4d_{101,\alpha}=3d_{122,\delta}$    
			&86\\
			
			2&$[1\bar20]_{\delta}\;\big|\big|\;[111]_\alpha$ &semi-coherent &$3d_{101,\alpha}=2d_{003,\delta}$ &10\\
			
			3&$[\bar47\;10]_{\delta}\;\big|\big|\;[111]_\alpha$&semi-coherent&-&2\\
			
			
			
			4&$[\bar41\;\bar{11}]_{\delta}\;\big|\big|\;[111]_\alpha$ &semi-coherent &$d_{101,\alpha}\simeq d_{31\bar1,\delta}$&2\\\bottomrule
			
		\end{tabular}
	\end{center}
	\label{tab:ORs}
\end{table}

\begin{figure}[!hbt]
	\centering
	\subfigure[]{\includegraphics[width=0.36\linewidth]{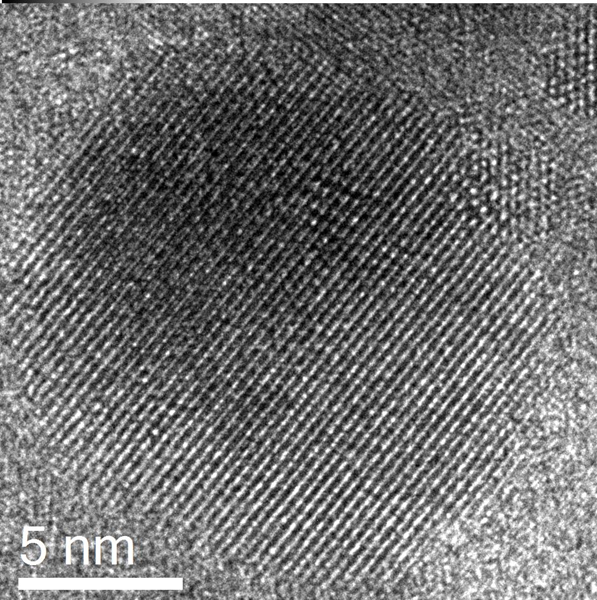}}
	\subfigure[]{\includegraphics[width=0.36\linewidth]{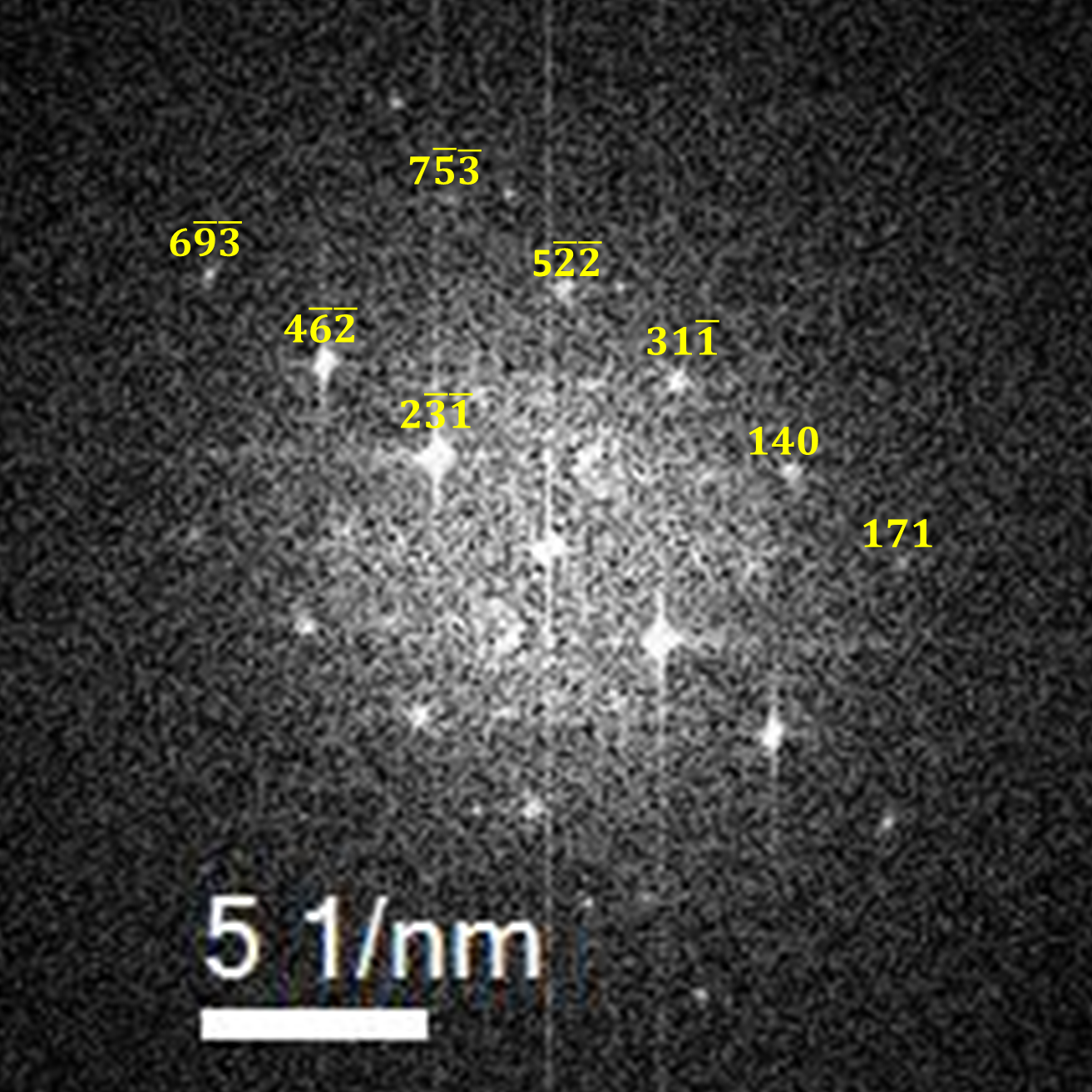}}
	\subfigure[]{\includegraphics[width=0.36\linewidth]{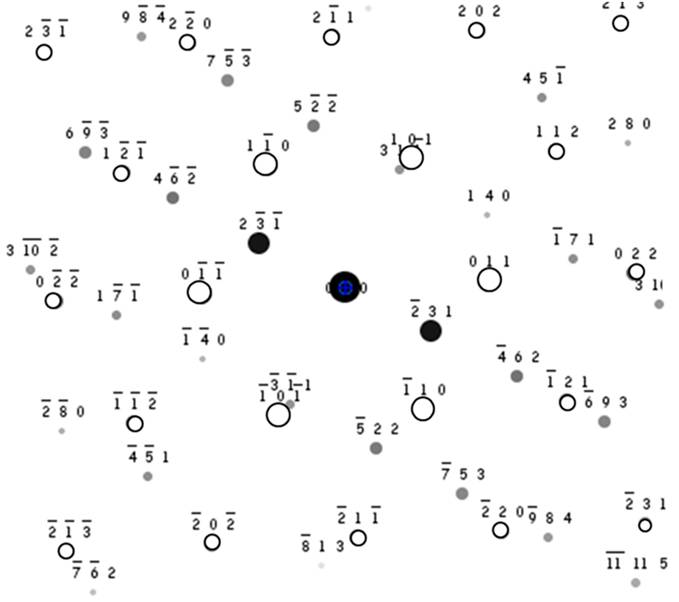}}
\subfigure[]{\includegraphics[width=0.36\linewidth]{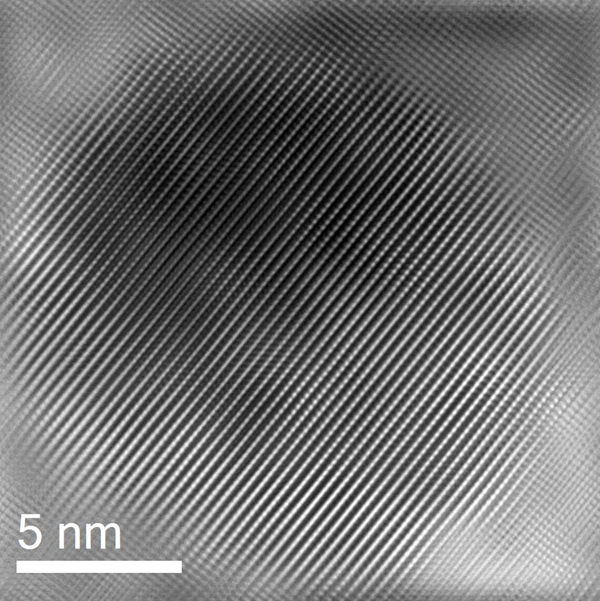}}
	\captionof{figure}{The HRTEM micrograph and its power spectrum is shown in (a) and (b). The spots marked in (b) correspond to [$\bar4$1 $\bar{11}]_\delta$. The simulated diffraction pattern and Bragg filtered image are  shown in (c) and (d) respectively.}
	\label{still024}
	\bigskip
	\ContinuedFloat
	\captionof{table}{Inter-planar distances (d) and angles ($\alpha$) of the nanoparticle of Figure \ref{still024}}
	\bigskip
	\begin{tabular}{cccccccc}
		\toprule
		\medskip
		&d(\AA),$\alpha(^\circ)$ &$d_1$ &$d_2$ &$d_3$ &$\alpha_{12}$ &$\alpha_{13}$ &$\alpha_{23}$\\\midrule
		&Measured &1.9  &1.59 &3.0 &56  &88  &32\\
		\medskip
		&Planes	&$\left(\bar42\bar3\right)$	&$\left(50\bar2\right)$	&$\left(1\bar21\right)$	&-	&-	&- \\ 
		\medskip
		&Calculated    &1.9 &1.58	&3.0 	&56	&88	&32\\\bottomrule
		\medskip
	\end{tabular}
	\label{tab:still024}
\end{figure}

\section{Prediction of  possible ORs in trigonal/bcc interface}
Different crystallographic models, edge to edge model, structural ledge model, near coincidence site model, invariant line model and O-lattice theory, have been successfully implemented in bcc/hcp and bcc/fcc interfaces to predict possible ORs and corresponding habit planes. Prediction of ORs is based on the fact that the minimization of strain energy at the interface is achieved by good atomic matching resulting from the parallelism between close packed planes or nearly close packed planes and close packed or nearly close packed directions of the two structures. So in order to predict the plausible ORs in current system, we start by considering close packed and nearly close packed planes and directions 
\par The planes with the largest value of structure factor, $|F_{hkl}|$, is the most closely packed planes in a crystal. With this criterion, the close packed and nearly close packed planes in bcc Fe is identified to be $\{110\}_\alpha$, $\{200\}_\alpha$ and $\{211\}_\alpha$ with corresponding d values $d_{110}=2.02$ {\AA}, $d_{200}=1.42$ {\AA}  and $d_{211}=1.16$ {\AA} respectively. 
Similarly, the structure factors of nearly 500 planes in Y\textsubscript{4}Zr\textsubscript{3}O\textsubscript{12} unitcell is calculated and six families of planes with considerably large structure factor values than other planes are identified as the most close packed and nearly close packed planes in that structure. They are: $\{211\}_\delta$, $\{003\}_\delta$, $\{2\bar10\}_\delta$, 
$\{21\bar2\}_\delta$, $\{012\}_\delta$ and $\{20\bar1\}_\delta$. The members in a family of planes are related by the symmetry of the crystal and are identical in all respects. However, the identification of individual members in a family of planes in $R\bar3$ space group is not straight forward as in crystals with cubic symmetry. Therefore, the members of each family of close packed and nearly close packed planes are calculated and listed in Table \ref{planes}. The d-spacings, structure factors and multiplicity of these family of planes are also included. With these family of planes and three close packed plane families in bcc, a total of eighteen probable trigonal/bcc plane pairs can be deduced. 
 \begin{table}[!hbt]
 	\ra{1.5}
 	\captionof{table}{The structure factors, d-spacings, multiplicity and individual members of the close packed and nearly close packed family of planes in Y\textsubscript{4}Zr\textsubscript{3}O\textsubscript{12}}
 	\begin{center}
 		\begin{tabular}{cp{2cm}p{2cm}p{2cm}p{2cm}p{4cm}}
 			\toprule
 			Sl.No.&Family {hkl} &Structure Factor ($|F_{hkl}|$) &d-spacing (\AA) &Multiplicity& Planes in the family \\\midrule
 			1& $\{211\}_\delta$&77.71 &3.00&6&$(\bar321)_\delta$, $(3\bar2\bar1)_\delta$, $(211)_\delta$, $(\bar2\bar1\bar1)_\delta$, $(1\bar31)_\delta$, $(\bar13\bar1)_\delta$  \\  			
 			2&$\{003\}_\delta$ &67.85   &3.03 &2&$(003)_\delta$, $(00\bar3)_\delta$ \\			
 			3&$\{2\bar10\}_\delta$&30.81 &4.86 &6& $(2\bar10)_\delta$, $(\bar210)_\delta$, $(110)_\delta$, $(\bar1\bar10)_\delta$, $(1\bar20)_\delta$, $(\bar120)_\delta$  \\					
 			4&$\{21\bar2\}_\delta$&24.23 &2.61 &6&$(3\bar22)_\delta$, $(\bar32\bar2)_\delta$, $(21\bar2)_\delta$, $(\bar2\bar12)_\delta$, $(1\bar3\bar2)_\delta$, $(\bar132)_\delta$\\
 			5&$\{012\}_\delta$ &23.62 &4.00 &6&$(10\bar2)_\delta$, $(\bar102)_\delta$, $(0\bar1\bar2)_\delta$,  $(012)_\delta$, $(\bar11\bar2)_\delta$, $(1\bar12)_\delta$  \\
 			6&$\{20\bar1\}_\delta$ &8.65    &3.82 &6&$(20\bar1)_\delta$, $(\bar201)_\delta$, $(021)_\delta$, $(0\bar2\bar1)_\delta$, $(2\bar21)_\delta$, $(\bar22\bar1)_\delta$\\
 			\bottomrule			
 		\end{tabular}
 	\end{center}
 	\label{planes}
 \end{table}

\par The next step is to identify close packed and nearly close packed directions. There are four close packed or nearly close packed directions in bcc. They are $<111>_\alpha$, $<110>_\alpha$, $<100>_\alpha$ and $<113>_\alpha$. The first three directions are straight directions and the last one is a zigzag direction\cite{zhang2005edge}. The atomic spacing in $<111>_\alpha$ direction is 2.47 {\AA}. The atomic spacing along $<100>_\alpha$ is equal to the lattice parameter ($a_\alpha$), 2.86 {\AA}. In $<110>_\alpha$ direction it is 4.04 {\AA} and in $<113>_\alpha$ direction it is $0.25a_\alpha\sqrt{11}$, which gives a value of 2.37 {\AA}\cite{zhang2005edge}. 
\par 
The close packed and nearly close packed directions in Y\textsubscript{4}Zr\textsubscript{3}O\textsubscript{12} are found and tabulated in Table \ref{dirn_family}. The individual members of the family of planes, generated by space group generators are also included. The most closely packed direction is $<1\bar20>_\delta$ and the interatomic spacing in this direction is 3.58 {\AA}, which is a straight atom row.  $<\bar111>_\delta$ and $<\bar510>_\delta$ are other straight directions  with interatomic spacings 6.37 {\AA} and 6.21 {\AA} respectively. 
\begin{table}[!hbt]
	\ra{1.3}
	\captionof{table}{Close packed and nearly close packed family of directions in Y\textsubscript{4}Zr\textsubscript{3}O\textsubscript{12}. The interatomic spacing in the direction, Multiplicity and Directions belonging to the family are also given.}
	\begin{center}
		\begin{tabular}{lp{2cm}p{2cm}p{2cm}p{5cm}}
			\toprule
			Sl.No.&Family &Interatomic Spacing ({\AA}) &Multiplicity&Directions belonging to the family \\\midrule
			1& $<1\bar20>_\delta$ & 3.58  &6&$[1\bar20]_\delta$, $[230]_\delta$, $[\bar3\bar10]_\delta$, $[\bar120]_\delta$, $[\bar2\bar30]_\delta$, $[310]_\delta$\\  			
			2&$<\bar111>_\delta$ &6.37 &6&$[\bar111]_\delta$, $[\bar1\bar21]_\delta$, $[12\bar1]_\delta$, $[211]_\delta$, $[1\bar1\bar1]_\delta$,$[\bar2\bar1\bar1]_\delta$   \\			
			3&$<\bar510>_\delta$&6.21&6&$[\bar510]_\delta$, $[5\bar10]_\delta$, $[650]_\delta$, $[\bar6\bar50]_\delta$, $[160]_\delta$, $[\bar1\bar60]_\delta$ \\					
		\bottomrule			
		\end{tabular}
	\end{center}
	\label{dirn_family}
\end{table}
\FloatBarrier
Therefore there can be nine directional combinations between trigonal Y\textsubscript{4}Zr\textsubscript{3}O\textsubscript{12} and bcc Fe, which are : $<1\bar20>_\delta||<111>_\alpha$, $<1\bar20>_\delta||<110>_\alpha$, $<1\bar20>_\delta||<100>_\alpha$,
$<\bar111>_\delta||<111>_\alpha$, $<\bar111>_\delta||<100>_\alpha$, $<\bar111>_\delta||<110>_\alpha$,  $<\bar510>_\delta||<111>_\alpha$, $<\bar510>_\delta||<100>_\alpha$ and $<\bar510>_\delta||<110>_\alpha$.  

\par For brevity we are considering combinations including only the close packed direction $<1\bar20>_\delta$ for further analysis. They are: $<1\bar20>_\delta||<111>_\alpha$, $<1\bar20>_\delta||<100>_\alpha$ and $<1\bar20>_\delta||<110>_\alpha$.
The probable ORs are determined by constructing combined stereographic projections corresponding to directional matches and
rotating them to match with close packed planes of bcc structure to find possible ORs with parallel directions and parallel plane combinations. The possible ORs are tabulated in Table \ref{ORs}.
\par In addition to the ORs shown in Figure \ref{sterogram}, three more close packed plane combinations are possible corresponding to $<1\bar20>_\delta||<111>_\alpha$ directional match, which give rise to three more ORs related by rotation of $\sim$ 5$^\circ$, as depicted in Figure \ref{sterogram2} (see Appendix A). Likewise, six rotationally related ORs can be derived for each of $<1\bar20>_\delta||<110>_\alpha$ and $<1\bar20>_\delta||<100>_\alpha$ directional combinations, as shown in Figure \ref{sterogram3} and Figure \ref{sterogram4} respectively of Appendix A.  This includes Y\textsubscript{4}Zr\textsubscript{3}O\textsubscript{12}/bcc-Fe orientation relationship observed by Dou \textit{et al.}\cite{dou2014tem,dou2020effects}. 
\begin{table}
	\ra{1.3}
	\captionof{table}{Summary of the possible orientation relationships between bcc Fe and trigonal Y\textsubscript{4}Zr\textsubscript{3}O\textsubscript{12}}
	\begin{center}
		\begin{tabular}{llp{3cm}}
			\toprule
			Sl. No. &OR &References\\\midrule
			1&$<1\bar20>_\delta||<111>_\alpha \text{ and } \{211\}_\delta||\{110\}_\alpha$&\\ 
			2&$<1\bar20>_\delta||<111>_\alpha \text{ and } \{003\}_\delta||\{110\}_\alpha$&Present study\\
			3&$<1\bar20>_\delta||<111>_\alpha \text{ and } \{21\bar2\}_\delta||\{110\}_\alpha$&Present study\\
			4&$<1\bar20>_\delta||<111>_\alpha \text{ and } \{211\}_\delta||\{211\}_\alpha$&\\ 
			5&$<1\bar20>_\delta||<111>_\alpha \text{ and } \{003\}_\delta||\{211\}_\alpha$&\\
			6&$<1\bar20>_\delta||<111>_\alpha \text{ and } \{21\bar2\}_\delta||\{211\}_\alpha$&\\
			\midrule
			7&$<1\bar20>_\delta||<100>_\alpha \text{ and } \{211\}_\delta||\{110\}_\alpha$&\\
			8&$<1\bar20>_\delta||<100>_\alpha\text{ and } \{003\}_\delta||\{110\}_\alpha$&\\
			9&$<1\bar20>_\delta||<100>_\alpha \text{ and } \{21\bar2\}_\delta||\{110\}_\alpha$&\cite{dou2014tem,dou2020effects}\\
			10&$<1\bar20>_\delta||<100>_\alpha \text{ and } \{211\}_\delta||\{002\}_\alpha$&\\
			11&$<1\bar20>_\delta||<100>_\alpha\text{ and } \{003\}_\delta||\{002\}_\alpha$&\\
			12&$<1\bar20>_\delta||<100>_\alpha \text{ and } \{21\bar2\}_\delta||\{002\}_\alpha$&\\
			\midrule
			13&$<1\bar20>_\delta||<110>_\alpha \text{ and } \{211\}_\delta||\{110\}_\alpha$&\\
			14&$<1\bar20>_\delta||<110>_\alpha \text{ and } \{003\}_\delta||\{110\}_\alpha$&\\
			15&$<1\bar20>_\delta||<110>_\alpha \text{ and } \{21\bar2\}_\delta||\{110\}_\alpha$&\\
			16&$<1\bar20>_\delta||<110>_\alpha \text{ and } \{211\}_\delta||\{002\}_\alpha$&\\
			17&$<1\bar20>_\delta||<110>_\alpha \text{ and } \{003\}_\delta||\{002\}_\alpha$&\\
			18&$<1\bar20>_\delta||<110>_\alpha \text{ and } \{21\bar2\}_\delta||\{002\}_\alpha$&\\\bottomrule
			\label{ORs}
		\end{tabular}
	\end{center}
\end{table}
\FloatBarrier
\section{Discussions}
 
\par The orientation relationships between Y\textsubscript{4}Zr\textsubscript{3}O\textsubscript{12}/bcc in ODS steels with Zr is pivotal to predict its precipitate formation/growth/dissolution mechanism,  dispersion strengthening effect and radiation response under various operational conditions. It is interesting to find that all the precipitates form semi-coherent interfaces, with at-least one precipitate plane parallel to a matrix plane (Table \ref{tab:ORs}). In smaller precipitates close packed planes and close packed directions of both crystal structures are aligned parallel and strain is accommodated by  arrays of misfit dislocations, while a few larger precipitates tend to deviate from these criteria, by aligning in a random high order zone axes, without periodic misfit dislocations. This explicit precipitate-matrix correlation in atomic scale is a conclusive evidence to the fact that the Y\textsubscript{2}O\textsubscript{3} is dissolving during the milling and it is re-precipitating as Y-Zr-O nuclei during hot-consolidation in an energetically preferred manner. The low formation energy favours Y\textsubscript{4}Zr\textsubscript{3}O\textsubscript{12} than other structures in Y-Zr-O system\cite{mohan2018effect,mohan2020ab}. Existence of a few random orientations with high order zone axes  might be due to:(i) loss of coherency in the growth phase as the strain energy overweighs the interfacial energy\cite{hsiung2011hrtem} or (ii) they nucleate with various orientation relationships owing to the non-equilibrium nature of the ball milling process\cite{mao2015crystallographic}. Since the precipitate diameter of these ORs belong exclusively to larger side of the size distribution, the former reason seems feasible for the occurrence of ORs involving high order zone axes in the current study.
\par It is well known that the strengthening of materials with fine dispersions can occur either by Orowan bypassing mechanism or by precipitate shearing mechanism, depending on the extent of coherency between the precipitate and matrix\cite{dieter1986mechanical}. 
By empirical methods, it is possible to formulate a quantitative insight into the operative strengthening mechanism by the predominant orienatation relationship in this study. Our calculations show that (see Appendix B) Orowan bypassing mechanism is the predominant strengthening mechanism in Zr-ODS steel, which is in line with reported calculations in other ODS steels\cite{shen2016microstructural}. The observed strain field surrounding the precipitate will help to enhance the Orowan mechanism by increasing the effective barrier for dislocations and will also affect the migration of irradiation induced defects\cite{brailsford1981effect,mao2015crystallographic}.
\par The well known criteria for energetically favoured interfaces, parallelism between close packed or nearly close packed directions of matrix and precipitate is verified in a large number of bcc/fcc and hcp/bcc systems. The close connection of this crystallographic formulation to experimental data of this study imply that the principles of existing crystallographic models are extendable to interfaces involving crystals other than cubic or hexagonal structures. Consequently, experimental observation of a single OR will lead to the possibility of a set of rotationally related ORs.  
\par On comparing Table \ref{tab:ORs} and Table \ref{ORs}, it can also be deduced that the spacing between misfit dislocation doublets ($D_d$) is an important factor in determining the favorable orientation relationship in a system. The most preferred OR has a spacing between dislocation doublet of 0.9 nm and the ORs with dislocation spacing 0.6 nm are less preferred than that, eventhough there is perfect coincidence on the coinciding planes. The larger spacing indicate more number of coinciding planes in the coherent patch and an effective reduction of interfacial energy. Among the closely packed and nearly closely packed planes, the plan combinations with minimum frequency of misfit dislocations (maximum value for $D_d=\frac{d_1d_2}{|d_1-d_2|}$) are the most probable ones. This explains why ORs with $\{110\}_\alpha/\{2\bar10\}_\delta$ ($D_d=0.3$ nm), $\{110\}_\alpha/\{012\}_\delta$ ($D_d=0.4$ nm) and $\{110\}_\alpha/\{20\bar1\}_\delta$ ($D_d=0.4$ nm) plane combinations are consistently absent. This criteria also implies that due to the unfavorable interfacial energy arising by closely packed misfit dislocations, any plane combinations involving $\{211\}_\alpha$ and  close packed or nearly close packed planes of $\delta$ are least probable, though the angle of rotation connecting them are similar to that of plane combinations involving $\{110\}_\alpha$. The exceptional similarity of experimental data and  stereographic-projection-method put forward in this study suggests that this method can be successfully applied to find most probable orientation relationships in different crystal systems by finding the close packed and nearly close packed planes/directions and filtering by the minimum $D_d$ criteria.

\section{Conclusions}
Orientation relationships between trigonal Y\textsubscript{4}Zr\textsubscript{3}O\textsubscript{12} precipitate ($\delta$) and bcc Fe matrix ($\alpha$) in oxide dispersion strengthened (ODS) ferritic steel containing Y\textsubscript{2}O\textsubscript{3} and Zr, synthesized by ball milling and hot extrusion are determined using transmission electron microscopy and image simulations.  Nearly 86\% of the precipitates possess orientation relationship (OR1) defined by: $[1\bar20]_\delta\big|\big|[111]_\alpha$ and $(21\bar2)_\delta\big|\big|(\bar110)_\alpha$. A relative rotation of this predominant OR by 5 $^\circ$ brings another set of planes to coincidence and forms a new OR (OR2): $[1\bar20]_\delta\big|\big|[111]_\alpha$ and $(003)_\delta\big|\big|(\bar110)_\alpha$, which accounts for 10\% of the precipitates analyzed in this study. These ORs form semi-coherent precipitate- matrix interfaces with periodic array of misfit dislocation doublets and the rotational relationship between these two ORs are similar to the relationship between Burger's and Pitsche-Shrader ORs in bcc/hcp systems. Remaining 4\% of the precipitates have orientations with higher order zone axes given by: $[\bar47\;10]_\delta\big|\big|[111]_\alpha$ and $(12\bar1)_\delta\big|\big|(\bar110)_\alpha$ and $[\bar41\;\bar{11}]_\delta\big|\big|[111]_\alpha$ and $(31\bar1)_\delta\big|\big|(\bar110)_\alpha$.

The existence of predominant OR is conclusive of the fact that the Y\textsubscript{2}O\textsubscript{3} powder gets completely dissolved in the matrix during milling and re-precipitates as Y\textsubscript{4}Zr\textsubscript{3}O\textsubscript{12} in an energetically preferred direction during the hot-consolidation process.  Further,combined stereographic projections corresponding to the close packed directions of the crystal systems are constructed and successfully implemented in predicting the energetically favoured ORs of Y\textsubscript{4}Zr\textsubscript{3}O\textsubscript{12}/bcc system by matching close packed and nearly close packed planes. The energetically favoured orientation relationships predicted based on this model is consistant with the predominant ORs observed in the current study and other reported ORs in the system. On comparing the experimental data and the crystallographic formulation, the probability of occurance of an OR is determined by the possible interfacial misfit dislocation frequency, the interface with less frequent misfit dislocations having more probability and vice versa.

\section*{Acknowledgment}
\noindent The authors gratefully acknowledge Dr. S. Amirthapandian, MSG, IGCAR for TEM image acquisition.
\newpage

\section*{Methods}
\subsection*{Synthesis of ODS steel}
To manufacture an ODS steel with composition Fe-14Cr-0.6Zr-0.3Y\textsubscript{2}O\textsubscript{3} ( Zr-ODS), the alloy powders are produced by milling the elemental powders (99.9\% pure, \textit{Alfa Aesar}) and Y\textsubscript{2}O\textsubscript{3} in a high energy horizontal ball mill (Simoloyer CM08 mill) for 6h under argon atmosphere using stainless steel vial and hardened steel balls (diameter=5mm).  The ball to powder ratio is 10:1. In this process, the powder particles are successively deformed, fractured and cold-welded. 
\par The milled powders are filled in mild steel cans, degassed at 450$^\circ$C, and sealed. The sealed cans are then forged at 1050$^\circ$C and subsequently hot extruded at 1150$^\circ$C with an extrusion ratio of 9 to get rods of diameter 16 mm. The extruded rods are further annealed at 900$^\circ$C for 1h and then water quenched. The nominal composition of the as-prepared alloy is given in Table \ref{composition}. 
\begin{table}[!hbt]
	\ra{1.3}
	\caption{Chemical composition of the studied alloy obtained by ICP-AES, *ICP-MS and **titrimetry}
	\begin{center}
		\begin{tabular}{ccccc}
			\toprule	
			Fe &**Cr &*Y\textsubscript{2}O\textsubscript{3} &Zr&Al\\ 
			\midrule
			Balance &13.6&0.30&0.65&$<0.04$\\\bottomrule
		\end{tabular}
		\label{composition}
	\end{center}
\end{table}
\subsection*{Transmission Electron Microscopy}
TEM specimens were prepared twin jet electropolishing method and carbon extraction replica method. Discs of thickness 300 $\mu$m sliced from the extruded rods were thinned down to $\sim$30$\mu m$ and polished using the silicon carbide papers of successively finer grades (1000, 1200, 2400, 4000) and diamond polishing. Small discs of diameter 3mm were punched out of the polished specimens. The discs were then electropolished to perforation using an electrolyte of perchloric acid and ethanol in a 1:9 ratio at -35$^\circ$C and 25 V. 
\par The TEM samples are examined using the LIBRA 200FE (Carl Zeiss) transmission electron microscope operating at 200 kV 
and the micrographs are thoroughly analyzed using softwares Digital Micrograph\cite{schaffer2016digital} and ProcessDiffraction\cite{labar2008electron,labar2009electron,labar2005consistent}. The power spectra of micrographs were compared to diffraction patterns simulated using JEMS software\cite{stadelmann2012jems} for further confirmation. Visualizations of planes, directions and crystal structures are done using VESTA\cite{momma2008vesta}. 
\FloatBarrier
\subsection*{APPENDIX A: Stereographic projections describing ORs derivable from ORs 1,2 and 3 of Table \ref{ORs}}
\begin{figure}[!hbt]
	\centering
	\includegraphics[width=\textwidth]{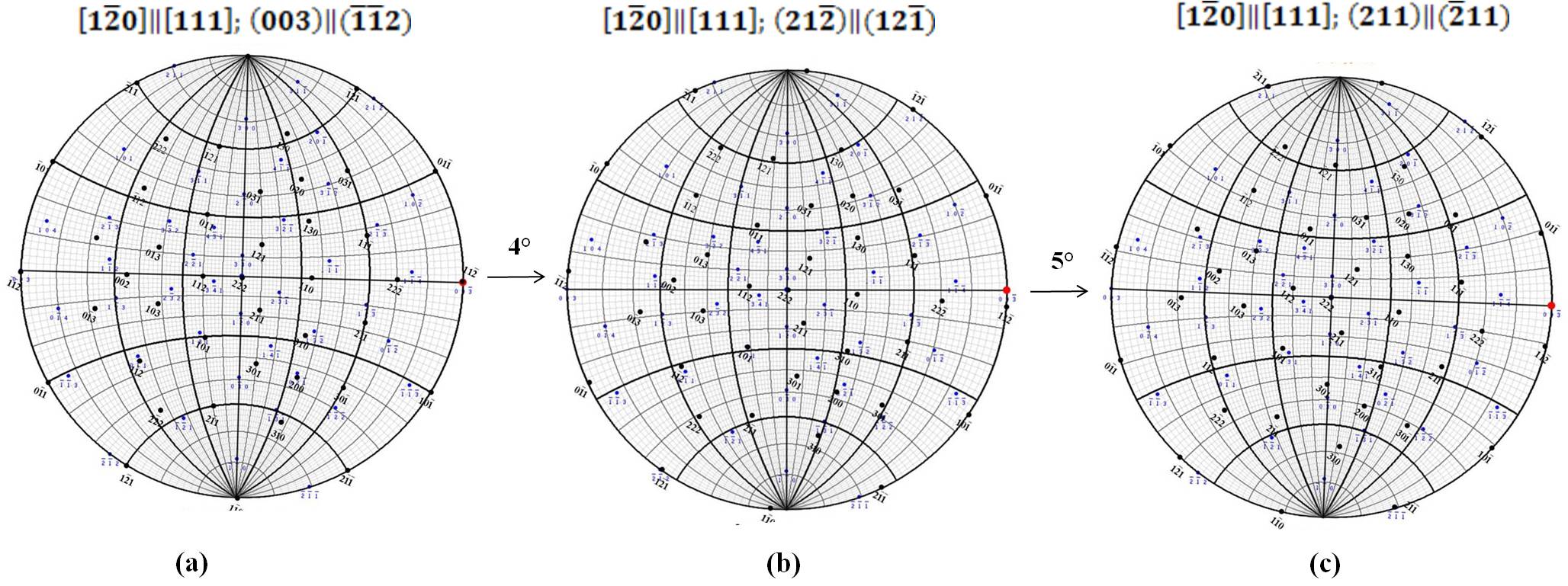}
	\caption{Combined stereographic projection of $[1\bar20]_\delta$ and $[111]_\alpha$}
	\label{sterogram2}
\end{figure}
\begin{figure}[!hbt]
	\centering
	\includegraphics[width=\textwidth]{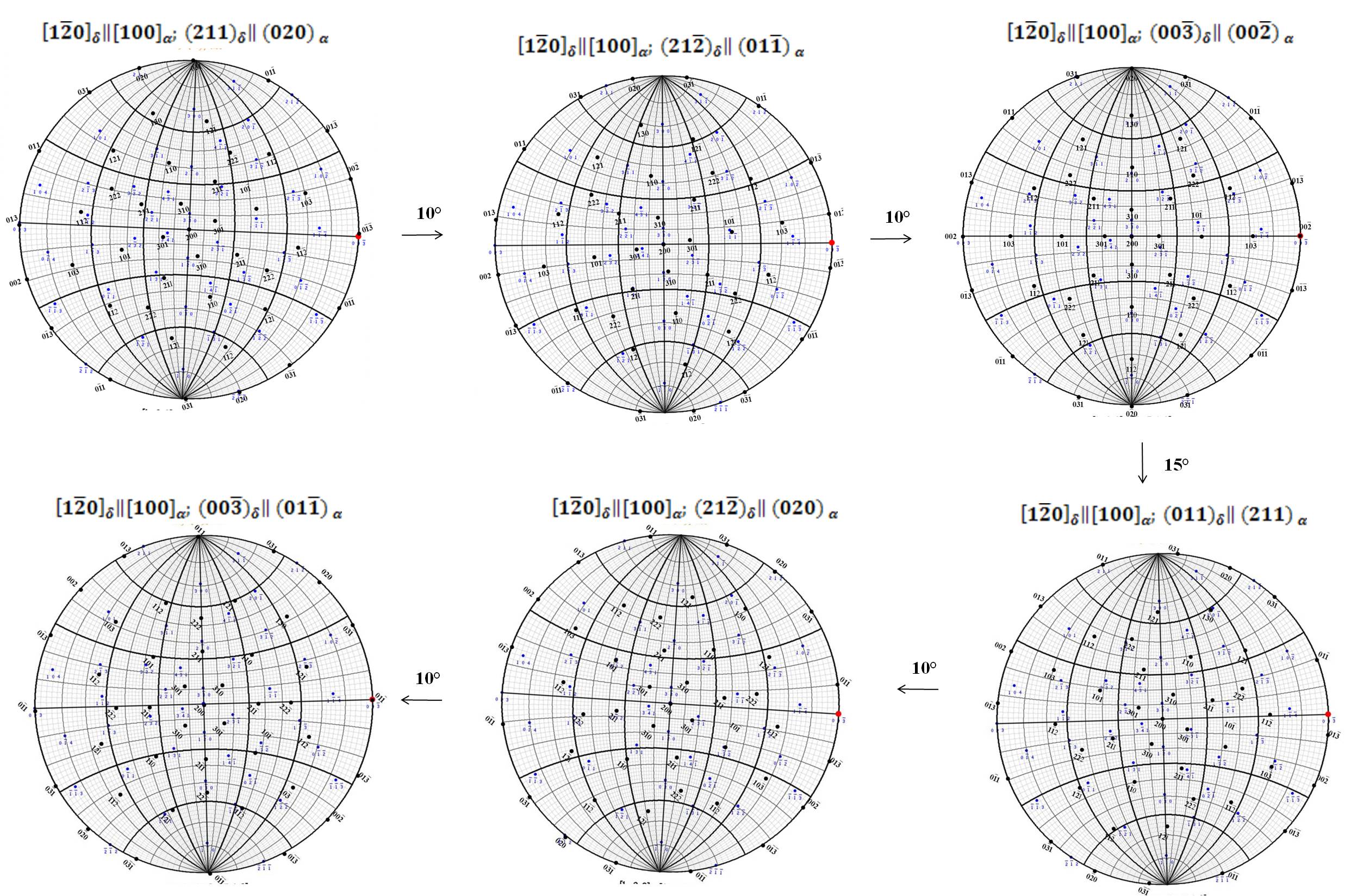}
	\caption{Combined stereographic projection of $[1\bar20]_\delta$ and $[100]_\alpha$ representing six plausible ORs generated by rotations. Blue spots indicate $[1\bar20]$ zone axis of Y\textsubscript{4}Zr\textsubscript{3}O\textsubscript{12} and black spots correspond to [111] zone axis of Fe}
	\label{sterogram3}
\end{figure}
\begin{figure}[!hbt]
	\centering
	\includegraphics[width=\textwidth]{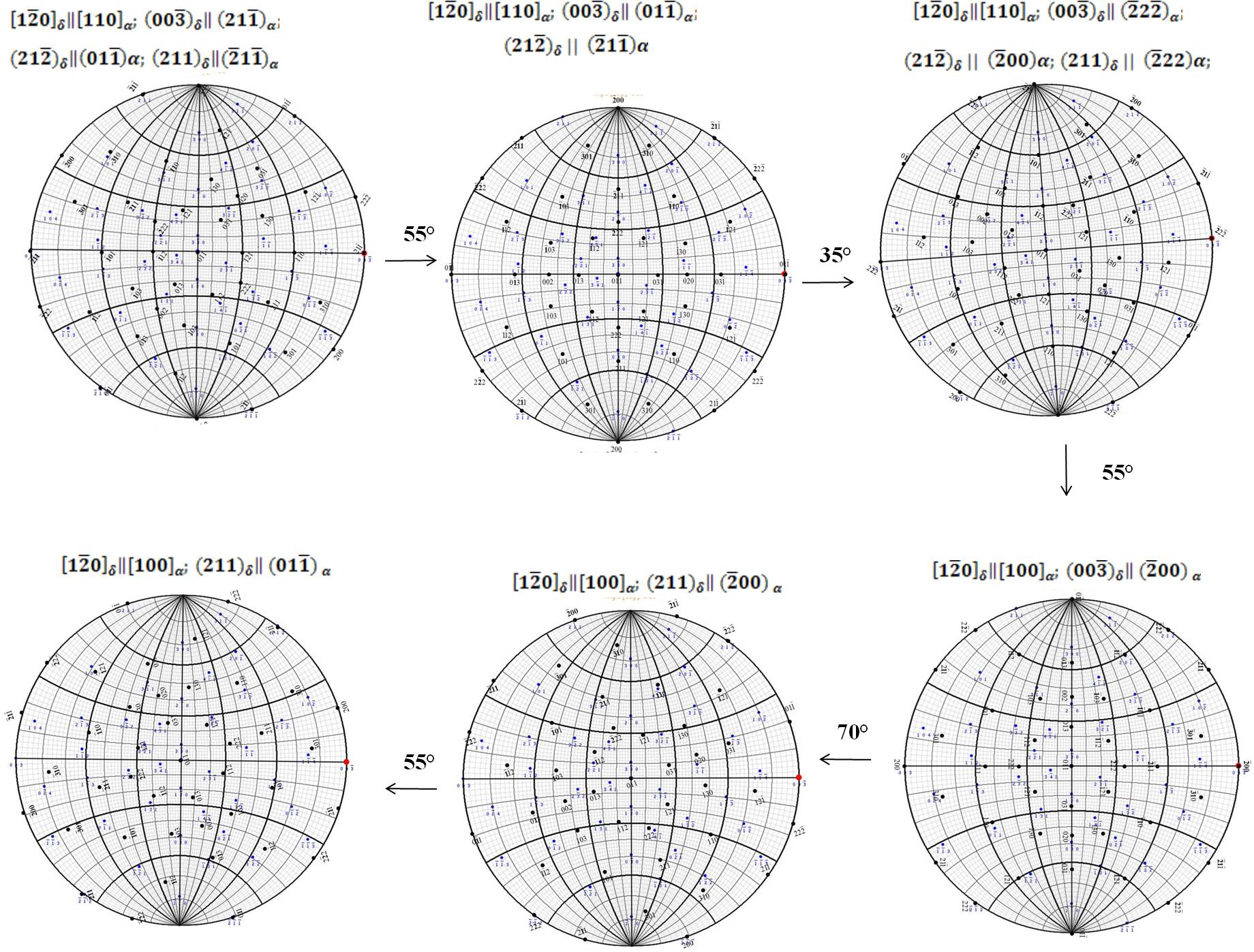}
	\caption{Combined stereographic projection of $[1\bar20]_\delta$ and $[110]_\alpha$ indicating six probable ORs. Blue spots indicate $[1\bar20]$ zone axis of Y\textsubscript{4}Zr\textsubscript{3}O\textsubscript{12} and black spots correspond to [111] zone axis of Fe}
	\label{sterogram4}
\end{figure}
\FloatBarrier
\newpage
\subsection*{APPENDIX B: Contribution of different strengthening mechanisms in Z\MakeLowercase{r}-ODS steel}

\par For the precipitates which are incoherent with the matrix, the dispersoid strengthening mechanism is governed by Orowan dislocation bypassing mechanism. The increment in yield strength is \cite{zhang2020evolution}:
\begin{equation}
\Delta\sigma_{orowan}=M\frac{0.4Gb}{\pi \sqrt{1-\nu}}\frac{ln(2\bar{r}/b)}{\lambda}; \bar{r}=\sqrt{\frac{2}{3}}r
\end{equation}
The meaning of parameters of this equation is summarized in Table \ref{parameters}.
\begin{table}[!hbt]
	\ra{1.3}
	\caption{Microstructural information of Zr-ODS}
	\begin{center}
		\begin{tabular}{lp{3 cm}p{4 cm}p{2 cm}}
			\toprule
			Symbol &Parameter &Value &Unit \\ \midrule
			
			M &Mean orientation factor &3.06, for polycrystalline, texture-free bcc metals\cite{keh1965work,kamikawa2015stress} &Dimensionless          \\ 
			
			G &Shear modulus &82\cite{kaye1911tables,kamikawa2015stress} &GPa  \\ 
			
			b &Magnitude of the Burger's vector &0.25\cite{ijiri2016oxide} &nm\\
			$\nu$ &Poissons Ratio &0.3\cite{kaye1911tables}  &Dimensionless\\
			
			$\alpha_\epsilon$  &a constant &0.2  &Dimensionless \\\bottomrule
		\end{tabular}
		\newline
		\label{parameters}
	\end{center}
\end{table}

The yield strength increment by Orowan bypass mechanism in Zr-ODS is $\sim$ 424 MPa

\par In the case of semi coherent precipitates, the strengthening mechanism can be Orowan bypassing or dislocation shearing mechanism. To determine the operative mechanism, the strength increment resulting from dislocation shearing has to be estimated and compared with $\Delta\sigma_{orowan}$ to find the operative mechanism. There are three factors contributing to the increase in yield strength by dislocation shearing : coherency strengthening($\Delta\sigma_{CS}$), modulus mismatch strengthening($\Delta\sigma_{MS}$) and order strengthening($\Delta\sigma_{OS}$). The equations available to calculate these strength increments are\cite{ma2014mechanical}:
\begin{align}
\Delta\sigma_{CS}&=M\times\alpha_\epsilon\times(G\epsilon_c)^{\frac{3}{2}}\times\sqrt{\frac{rf}{0.5Gb}}\\
\Delta\sigma_{MS}&=M\times0.0055(\Delta G)^{\frac{3}{2}}\times\sqrt{\frac{2f}{G}}\times \Big(\frac{r}{b}\Big)^{\frac{3m}{2}-1}\\
\Delta\sigma_{OS}&=M\times0.81\times\frac{\gamma_{apb}}{2b}\times\sqrt{\frac{3\pi f}{8}}
\end{align}

The larger of $(\Delta\sigma_{CS}+\Delta\sigma_{MS})$ or $\Delta\sigma_{OS}$ is the total strength increment from shearing mechanism.
\par 
The constrained effective coherent misfit strain, $\epsilon_c$ is defined as\cite{wang2013prediction}:
\begin{equation}
\epsilon_c=\frac{1}{3}\times\Big(\frac{1+\nu}{1-\nu}\Big)\times\Big(\frac{\epsilon_{eff}}{1+\frac{4G}{3B_c}}\Big)
\end{equation}

where, $B_c$ is the bulk modulus of the precipitate. $\epsilon_{eff}$ is the effective scalar coherency strain:
\begin{align}
\epsilon_{eff}=\frac{\sqrt2}{3}\Big[(\epsilon_{11}-\epsilon_{22})^2+(\epsilon_{22}-\epsilon_{33})^2+(\epsilon_{33}-\epsilon_{1})^2\Big]^{\frac{1}{2}}
\end{align}
where 
\begin{equation}
\epsilon_{11}=\frac{a_p}{a_m}-1;\quad \epsilon_{22}=\frac{\sqrt3a_p}{\sqrt2a_m}-1;\quad \epsilon_{33}=\frac{c_p}{\sqrt2a_m}-1
\end{equation}
$a_m$ is the lattice parameter of the Fe- matrix, 2.869\AA. $a_p$ and $c_p$ are the lattice parameters of the precipitate.
\par The calculated strength component due to coherency strengthening($\Delta\sigma_{CS}$) is $\sim$ 558 MPa for Zr-ODS and clearly $\Delta\sigma_{orowan} < \Delta\sigma_{CS}$. The component with smaller value is the predominant operating mechanism\cite{ma2014mechanical}. Therefore Orowan bypassing mechanism is the predominant strengthening mechanism in ODS steels. 
\bibliographystyle{lion-msc}
\bibliography{paper}

\begin{thebibliography}{10}

\bibitem{dahmen1982orientation}
U.~Dahmen, {\em Orientation relationships in precipitation systems},
\newblock Acta Metallurgica {\bf 30}, 63 (1982).

\bibitem{burgers1934process}
W.~Burgers, {\em On the process of transition of the cubic-body-centered
  modification into the hexagonal-close-packed modification of zirconium},
\newblock Physica {\bf 1}, 561 (1934).

\bibitem{rong1984crystallography}
W.~Rong and G.~Dunlop, {\em The crystallography of secondary carbide
  precipitation in high speed steel},
\newblock Acta Metallurgica {\bf 32}, 1591 (1984).

\bibitem{potter1973structure}
D.~Potter, {\em The structure, morphology and orientation relationship of
  {V$_3$N} in $\alpha$-vanadium},
\newblock Journal of the Less Common Metals {\bf 31}, 299 (1973).

\bibitem{pitsch1959martensite}
W.~Pitsch, {\em The martensite transformation in thin foils of iron-nitrogen
  alloys},
\newblock Philosophical Magazine {\bf 4}, 577 (1959).

\bibitem{ray1990transformation}
R.~Ray and J.~Jonas, {\em Transformation textures in steels},
\newblock International Materials Reviews {\bf 35}, 1 (1990).

\bibitem{kelly2006edge}
P.~Kelly and M.-X. Zhang, {\em Edge-to-edge matching—The fundamentals},
\newblock Metallurgical and Materials Transactions A {\bf 37}, 833 (2006).

\bibitem{zhang2005crystallographic}
M.-X. Zhang, P.~M. Kelly, M.~A. Easton, and J.~A. Taylor, {\em Crystallographic
  study of grain refinement in aluminum alloys using the edge-to-edge matching
  model},
\newblock Acta Materialia {\bf 53}, 1427 (2005).

\bibitem{zhang2005edge}
M.-X. Zhang and P.~Kelly, {\em Edge-to-edge matching and its applications: Part
  I. Application to the simple HCP/BCC system},
\newblock Acta Materialia {\bf 53}, 1073 (2005).

\bibitem{rigsbee1979computer}
J.~Rigsbee and H.~Aaronson, {\em A computer modeling study of partially
  coherent FCC: BCC boundaries},
\newblock Acta Metallurgica {\bf 27}, 351 (1979).

\bibitem{hall1972structure}
M.~Hall, H.~Aaronson, and K.~Kinsma, {\em The structure of nearly coherent fcc:
  bcc boundaries in a Cu Cr alloy},
\newblock Surface Science {\bf 31}, 257 (1972).

\bibitem{luo1987invariant}
C.~Luo and G.~Weatherly, {\em The invariant line and precipitation in a Ni-45
  wt\% Cr alloy},
\newblock Acta metallurgica {\bf 35}, 1963 (1987).

\bibitem{howe1992comparison}
J.~Howe and D.~Smith, {\em Comparison between the invariant line and structural
  ledge theories for predicting the habit plane, orientation relationship and
  interphase boundary structure of plate-shaped precipitates},
\newblock Acta metallurgica et materialia {\bf 40}, 2343 (1992).

\bibitem{dahmen1984invariant}
U.~Dahmen, P.~Ferguson, and K.~Westmacott, {\em Invariant line strain and
  needle-precipitate growth directions in Fe-Cu},
\newblock Acta Metallurgica {\bf 32}, 803 (1984).

\bibitem{balluffi1982csl}
R.~Balluffi, A.~Brokman, and A.~King, {\em CSL/DSC lattice model for general
  crystalcrystal boundaries and their line defects},
\newblock Acta Metallurgica {\bf 30}, 1453 (1982).

\bibitem{ribis2012interfacial}
J.~Ribis and Y.~De~Carlan, {\em Interfacial strained structure and orientation
  relationships of the nanosized oxide particles deduced from elasticity-driven
  morphology in oxide dispersion strengthened materials},
\newblock Acta materialia {\bf 60}, 238 (2012).

\bibitem{klimiankou2004tem}
M.~Klimiankou, R.~Lindau, and A.~M{\"o}slang, {\em TEM characterization of
  structure and composition of nanosized ODS particles in reduced activation
  ferritic--martensitic steels},
\newblock Journal of Nuclear Materials {\bf 329}, 347 (2004).

\bibitem{oka2013morphology}
H.~Oka, M.~Watanabe, N.~Hashimoto, S.~Ohnuki, S.~Yamashita, and S.~Ohtsuka,
  {\em Morphology of oxide particles in ODS austenitic stainless steel},
\newblock Journal of Nuclear Materials {\bf 442}, S164 (2013).

\bibitem{hirata2011atomic}
A.~Hirata, T.~Fujita, Y.~Wen, J.~Schneibel, C.~T. Liu, and M.~Chen, {\em Atomic
  structure of nanoclusters in oxide-dispersion-strengthened steels},
\newblock Nature materials {\bf 10}, 922 (2011).

\bibitem{hsiung2010formation}
L.~L. Hsiung, M.~J. Fluss, S.~J. Tumey, B.~W. Choi, Y.~Serruys, F.~Willaime,
  and A.~Kimura, {\em Formation mechanism and the role of nanoparticles in
  {Fe-Cr ODS} steels developed for radiation tolerance},
\newblock Physical Review B {\bf 82}, 184103 (2010).

\bibitem{dou2019morphology}
P.~Dou, W.~Sang, and A.~Kimura, {\em Morphology, crystal and metal/oxide
  interface structures of nanoparticles in Fe--15Cr--2W--0.5 Ti--7Al--0.4
  Zr--0.5 Y2O3 ODS steel},
\newblock Journal of Nuclear Materials {\bf 523}, 231 (2019).

\bibitem{dou2019crystal}
P.~Dou, L.~Qiu, S.~Jiang, and A.~Kimura, {\em Crystal and metal/oxide interface
  structures of nanoparticles in Fe--16Cr--0.1 Ti--0.35 Y2O3 ODS steel},
\newblock Journal of Nuclear Materials {\bf 523}, 320 (2019).

\bibitem{dou2014tem}
P.~Dou, A.~Kimura, R.~Kasada, T.~Okuda, M.~Inoue, S.~Ukai, S.~Ohnuki,
  T.~Fujisawa, and F.~Abe, {\em {TEM} and {HRTEM} study of oxide particles in
  an {Al-alloyed high-Cr} oxide dispersion strengthened steel with {Zr}
  addition},
\newblock Journal of Nuclear Materials {\bf 444}, 441 (2014).

\bibitem{odette2008recent}
G.~Odette, M.~Alinger, and B.~Wirth, {\em Recent developments in
  irradiation-resistant steels},
\newblock Annu. Rev. Mater. Res. {\bf 38}, 471 (2008).

\bibitem{alinger2004development}
M.~Alinger, G.~Odette, and D.~Hoelzer, {\em The development and stability of
  {Y--Ti--O} nanoclusters in mechanically alloyed {Fe--Cr} based ferritic
  alloys},
\newblock Journal of Nuclear Materials {\bf 329}, 382 (2004).

\bibitem{alinger2008role}
M.~Alinger, G.~Odette, and D.~Hoelzer, {\em On the role of alloy composition
  and processing variables in nanofeature formation and dispersion
  strengthening in nanostructured ferritic alloys},
\newblock Fusion Mater. Semiannu. Prog. Rep. DOE-ER-0313 {\bf 43} (2008).

\bibitem{ukai1993alloying}
S.~Ukai, M.~Harada, H.~Okada, M.~Inoue, S.~Nomura, S.~Shikakura, K.~Asabe,
  T.~Nishida, and M.~Fujiwara, {\em Alloying design of oxide dispersion
  strengthened ferritic steel for long life {FBRs} core materials},
\newblock Journal of Nuclear Materials {\bf 204}, 65 (1993).

\bibitem{mohan2020ab}
S.~Mohan, G.~Kaur, C.~David, B.~Panigrahi, and G.~Amarendra, {\em Ab initio
  molecular dynamics simulation of threshold displacement energies and defect
  formation energies in Y4Zr3O12},
\newblock Journal of Applied Physics {\bf 127}, 235901 (2020).

\bibitem{red1991crystal}
V.~Red'ko and L.~Lopato, {\em Crystal structure of the compounds
  {M$_4$Zr$_3$O$_{12}$} and {M$_4$Hf$_3$O$_{12}$}(where {M} is a rare earth
  element)},
\newblock Inorganic Materials {\bf 27}, 1609 (1991).

\bibitem{hsiung2011hrtem}
L.~Hsiung, M.~Fluss, S.~Tumey, J.~Kuntz, B.~El-Dasher, M.~Wall, B.~Choi,
  A.~Kimura, F.~Willaime, and Y.~Serruys, {\em {HRTEM} study of oxide
  nanoparticles in {K3-ODS} ferritic steel developed for radiation tolerance},
\newblock Journal of Nuclear Materials {\bf 409}, 72 (2011).

\bibitem{dong2017enhancement}
H.~Dong, L.~Yu, Y.~Liu, C.~Liu, H.~Li, and J.~Wu, {\em Enhancement of tensile
  properties due to microstructure optimization in ODS steels by zirconium
  addition},
\newblock Fusion Engineering and Design {\bf 125}, 402 (2017).

\bibitem{xu2017microstructure}
H.~Xu, Z.~Lu, D.~Wang, and C.~Liu, {\em Microstructure refinement and
  strengthening mechanisms of a {9Cr} oxide dispersion strengthened steel by
  zirconium addition},
\newblock Nuclear Engineering and Technology {\bf 49}, 178 (2017).

\bibitem{dou2020effects}
P.~Dou, S.~Jiang, L.~Qiu, and A.~Kimura, {\em Effects of contents of {Al, Zr
  and Ti} on oxide particles in {Fe--15Cr--2W--0.35 Y$_2$O$_3$} {ODS} steels},
\newblock Journal of Nuclear Materials {\bf 531}, 152025 (2020).

\bibitem{ribis2013influence}
J.~Ribis, M.-L. Lescoat, S.~Zhong, M.-H. Mathon, and Y.~De~Carlan, {\em
  Influence of the low interfacial density energy on the coarsening resistivity
  of the nano-oxide particles in Ti-added ODS material},
\newblock Journal of Nuclear Materials {\bf 442}, S101 (2013).

\bibitem{carter2016transmission}
C.~B. Carter and D.~B. Williams,
\newblock {\em Transmission electron microscopy: Diffraction, imaging, and
  spectrometry},
\newblock Springer, 2016.

\bibitem{matsukawa2019crystallography}
Y.~Matsukawa,
\newblock {\em Crystallography of Precipitates in Metals and Alloys:(1)
  {A}nalysis of Crystallography},
\newblock in {\em Crystallography}, IntechOpen, 2019.

\bibitem{mohan2018effect}
S.~Mohan, G.~Kaur, B.~Panigrahi, C.~David, and G.~Amarendra, {\em Effect of Zr
  and Al addition on nanocluster formation in oxide dispersion strengthened
  steel-an ab initio study},
\newblock Journal of Alloys and Compounds {\bf 767}, 122 (2018).

\bibitem{mao2015crystallographic}
X.~Mao, T.~K. Kim, S.~S. Kim, Y.~S. Han, K.~H. Oh, and J.~Jang, {\em
  Crystallographic relationship of YTaO4 particles with matrix in Ta-containing
  12Cr ODS steel},
\newblock Journal of Nuclear Materials {\bf 461}, 329 (2015).

\bibitem{dieter1986mechanical}
G.~E. Dieter and D.~J. Bacon,
\newblock {\em Mechanical {Metallurgy}}, volume~3,
\newblock McGraw-hill New York, 1986.

\bibitem{shen2016microstructural}
J.~Shen, Y.~Li, F.~Li, H.~Yang, Z.~Zhao, S.~Kano, Y.~Matsukawa, Y.~Satoh, and
  H.~Abe, {\em Microstructural characterization and strengthening mechanisms of
  a 12Cr-ODS steel},
\newblock Materials Science and Engineering: A {\bf 673}, 624 (2016).

\bibitem{brailsford1981effect}
A.~Brailsford and L.~Mansur, {\em The effect of precipitate-matrix interface
  sinks on the growth of voids in the matrix},
\newblock Journal of Nuclear Materials {\bf 104}, 1403 (1981).

\bibitem{schaffer2016digital}
B.~Schaffer,
\newblock {\em Digital {M}icrograph},
\newblock in {\em Transmission Electron Microscopy}, pages 167--196, Springer,
  2016.

\bibitem{labar2008electron}
J.~L. L{\'a}b{\'a}r, {\em Electron diffraction based analysis of phase
  fractions and texture in nanocrystalline thin films, Part {I}: {P}rinciples},
\newblock Microscopy and Microanalysis {\bf 14}, 287 (2008).

\bibitem{labar2009electron}
J.~L. L{\'a}b{\'a}r, {\em Electron diffraction based analysis of phase
  fractions and texture in nanocrystalline thin films, Part {II}:
  {I}mplementation},
\newblock Microscopy and Microanalysis {\bf 15}, 20 (2009).

\bibitem{labar2005consistent}
J.~L. L{\'a}b{\'a}r, {\em Consistent indexing of a (set of) single crystal
  {SAED} pattern (s) with the {ProcessDiffraction} program},
\newblock Ultramicroscopy {\bf 103}, 237 (2005).

\bibitem{stadelmann2012jems}
P.~Stadelmann, {\em {JEMS} electron microscopy software},
\newblock V4, EPFL, Lausanne, Switzerland  (2012).

\bibitem{momma2008vesta}
K.~Momma and F.~Izumi, {\em {VESTA}: {A} three-dimensional visualization system
  for electronic and structural analysis},
\newblock Journal of Applied Crystallography {\bf 41}, 653 (2008).

\bibitem{zhang2020evolution}
W.~Zhang, Z.~Zhao, J.~Fang, P.~He, Z.~Chao, D.~Gong, G.~Chen, and L.~Jiang,
  {\em Evolution and strengthening mechanism of metastable precipitates in
  {Cu-2.0 wt\% Be} alloy},
\newblock Journal of Alloys and Compounds , 157601 (2020).

\bibitem{keh1965work}
A.~Keh, {\em Work hardening and deformation sub-structure in iron single
  crystals deformed in tension at {298 K}},
\newblock Philosophical Magazine {\bf 12}, 9 (1965).

\bibitem{kamikawa2015stress}
N.~Kamikawa, K.~Sato, G.~Miyamoto, M.~Murayama, N.~Sekido, K.~Tsuzaki, and
  T.~Furuhara, {\em Stress--strain behavior of ferrite and bainite with
  nano-precipitation in low carbon steels},
\newblock Acta Materialia {\bf 83}, 383 (2015).

\bibitem{kaye1911tables}
G.~W.~C. Kaye and T.~H. Laby,
\newblock {\em Tables of physical and chemical constants and some mathematical
  functions},
\newblock Longmans, Green and Company, 1911.

\bibitem{ijiri2016oxide}
Y.~Ijiri, N.~Oono, S.~Ukai, S.~Ohtsuka, T.~Kaito, and Y.~Matsukawa, {\em Oxide
  particle--dislocation interaction in {9Cr-ODS} steel},
\newblock Nuclear Materials and Energy {\bf 9}, 378 (2016).

\bibitem{ma2014mechanical}
K.~Ma, H.~Wen, T.~Hu, T.~D. Topping, D.~Isheim, D.~N. Seidman, E.~J. Lavernia,
  and J.~M. Schoenung, {\em Mechanical behavior and strengthening mechanisms in
  ultrafine grain precipitation-strengthened aluminum alloy},
\newblock Acta Materialia {\bf 62}, 141 (2014).

\bibitem{wang2013prediction}
J.-S. Wang, M.~Mulholland, G.~B. son, and D.~N. Seidman, {\em Prediction of the
  yield strength of a secondary-hardening steel},
\newblock Acta Materialia {\bf 61}, 4939 (2013).

\end{thebibliography}
\end{document}